\documentclass[graybox,natbib]{svmult}

\usepackage{mathptmx}       
\usepackage{helvet}         
\usepackage{courier}        
\usepackage{type1cm}        
\usepackage{makeidx}         
\usepackage{graphicx}        
\usepackage{multicol}        
\usepackage[bottom]{footmisc}
\usepackage{amsmath,amssymb,bm}
\usepackage[utf8]{inputenc}  
\newcommand{\bfr}{{\textbf{\itshape r}}} 
\newcommand{\bff}{\textbf{\itshape F}}
\newcommand{\bfg}{\textbf{\itshape g}}
\newcommand{\bfk}{\boldsymbol{\kappa}}
\newcommand{\bfv}{{\textbf{\itshape\ttfamily{v}}}}

\def\dd{{\rm d}}

\makeindex             

\begin{document}

\newcommand*\aap{A\&A}
\let\astap=\aap
\newcommand*\aapr{A\&A~Rev.}
\newcommand*\aaps{A\&AS}
\newcommand*\actaa{Acta Astron.}
\newcommand*\aj{AJ}
\newcommand*\ao{Appl.~Opt.}
\let\applopt\ao
\newcommand*\apj{ApJ}
\newcommand*\apjl{ApJ}
\let\apjlett\apjl
\newcommand*\apjs{ApJS}
\let\apjsupp\apjs
\newcommand*\aplett{Astrophys.~Lett.}
\newcommand*\apspr{Astrophys.~Space~Phys.~Res.}
\newcommand*\apss{Ap\&SS}
\newcommand*\araa{ARA\&A}
\newcommand*\azh{AZh}
\newcommand*\baas{BAAS}
\newcommand*\bac{Bull. astr. Inst. Czechosl.}
\newcommand*\bain{Bull.~Astron.~Inst.~Netherlands}
\newcommand*\caa{Chinese Astron. Astrophys.}
\newcommand*\cjaa{Chinese J. Astron. Astrophys.}
\newcommand*\fcp{Fund.~Cosmic~Phys.}
\newcommand*\gca{Geochim.~Cosmochim.~Acta}
\newcommand*\grl{Geophys.~Res.~Lett.}
\newcommand*\iaucirc{IAU~Circ.}
\newcommand*\icarus{Icarus}
\newcommand*\jcap{J. Cosmology Astropart. Phys.}
\newcommand*\jcp{J.~Chem.~Phys.}
\newcommand*\jgr{J.~Geophys.~Res.}
\newcommand*\jqsrt{J.~Quant.~Spec.~Radiat.~Transf.}
\newcommand*\jrasc{JRASC}
\newcommand*\memras{MmRAS}
\newcommand*\memsai{Mem.~Soc.~Astron.~Italiana}
\newcommand*\mnras{MNRAS}
\newcommand*\na{New A}
\newcommand*\nar{New A Rev.}
\newcommand*\nat{Nature}
\newcommand*\nphysa{Nucl.~Phys.~A}
\newcommand*\pasa{PASA}
\newcommand*\pasj{PASJ}
\newcommand*\pasp{PASP}
\newcommand*\physrep{Phys.~Rep.}
\newcommand*\physscr{Phys.~Scr}
\newcommand*\planss{Planet.~Space~Sci.}
\newcommand*\pra{Phys.~Rev.~A}
\newcommand*\prb{Phys.~Rev.~B}
\newcommand*\prc{Phys.~Rev.~C}
\newcommand*\prd{Phys.~Rev.~D}
\newcommand*\pre{Phys.~Rev.~E}
\newcommand*\prl{Phys.~Rev.~Lett.}
\newcommand*\procspie{Proc.~SPIE}
\newcommand*\qjras{QJRAS}
\newcommand*\rmxaa{Rev. Mexicana Astron. Astrofis.}
\newcommand*\skytel{S\&T}
\newcommand*\solphys{Sol.~Phys.}
\newcommand*\sovast{Soviet~Ast.}
\newcommand*\ssr{Space~Sci.~Rev.}
\newcommand*\zap{ZAp}

\title*{Theory of Stellar Oscillations}
\author{Margarida S.~Cunha}
\institute{Margarida S.~Cunha \at Instituto de Astrof\'isica e Ci\^encias do Espa\c{c}o, Universidade do Porto, CAUP, Rua das Estrelas, 4150-762 Porto, Portugal,\\ 
\email{mcunha@astro.up.pt}}
%
%
\maketitle


\abstract{In recent years, astronomers have witnessed
  major progresses in the field of stellar physics. This was made possible thanks to the
combination of a solid theoretical understanding of the
phenomena of stellar pulsations and the availability of a tremendous
amount of exquisite space-based asteroseismic data. In this context, this chapter reviews the basic theory of stellar pulsations,
considering small, adiabatic perturbations to a static, spherically
symmetric equilibrium. It starts with a brief discussion of the solar
oscillation spectrum, followed by the setting of the theoretical
problem, including the presentation of the equations of
hydrodynamics, their perturbation, and a discussion of the
functional form of the
solutions.  Emphasis is put on the physical properties of the
different types of modes, in particular acoustic (p-) and gravity
(g-) modes and their propagation cavities. The surface (f-) mode
solutions are also discussed. While not attempting to be 
comprehensive,  it is hoped that the summary presented in this chapter addresses
the most important theoretical aspects that are required for a solid start
in stellar pulsations research. }

\section{Introduction}
\label{intro}


The study of stellar pulsations is revolutionizing our knowledge
of the internal structure and dynamics of stars and, as a consequence, also our
understanding of stellar evolution. This is made possible
through the combination of a solid theoretical understanding of the
phenomena of stellar pulsations and the availability of a tremendous
amount of high-quality data, in particular that acquired from space
with satellites such as \textit{SOHO} \citep{domingo95}, observing the Sun for
over 20 years, and \textit{CoRoT} \citep{baglin06}
and \textit{Kepler} \citep{guilliland10,koch10}. 
 
In this chapter I review basic aspects of the theory of stellar
pulsations. Given the limited space available, options
had to be made on what to discuss. A more detailed view of the
aspects considered here, as well as discussions of the issues that have
been left out can be found, e.g., in
published books \citep{unno89,aerts10},
lecture notes \citep{gough93}, as well as in other long reviews \citep{cunhaetal07,basu16}.

\begin{figure}[t]
\includegraphics[scale=.65]{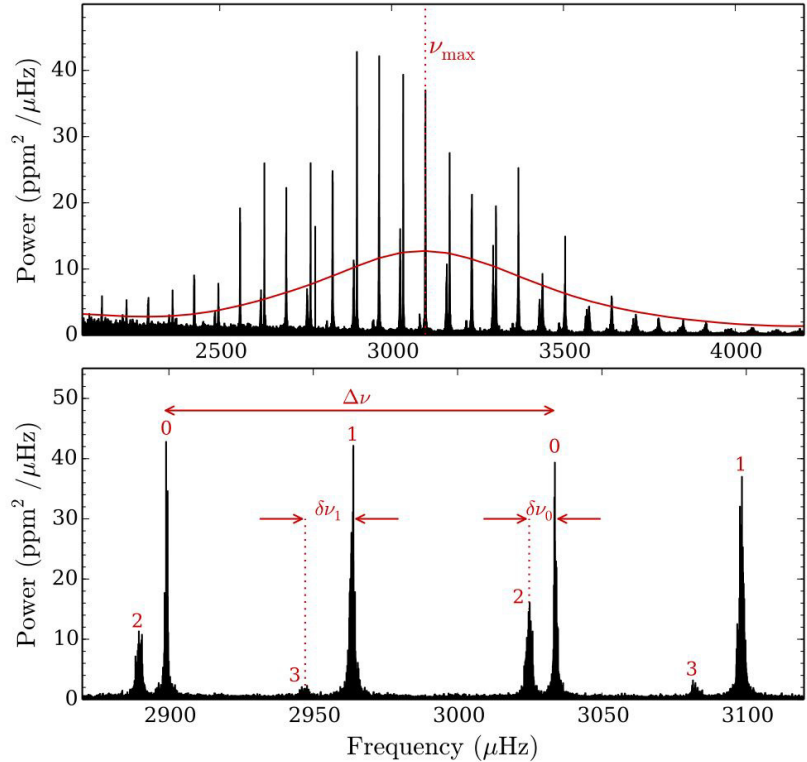}
%
%
\caption{Power density spectrum of the Sun obtained from data acquired with
  VIRGO/SPM 
  onboard the \textit{SOHO} satellite \citep{frohlich95,jimenez02}. \textit{Top:} The
  red line shows the  power spectrum density smoothed by
  $3\Delta\nu$ and multiplied by 50, used to estimate $\nu_{\rm max}$. \textit{Bottom:}
A zoom of the upper panel illustrating a few modes, identified
by mode degree. The large frequency separation, $\Delta\nu$, and the
small frequency separations between pairs of modes with degrees $l=0,
2$, $\delta\nu_0$, and pairs of modes with degrees $l=1,3$,
$\delta\nu_1$, are also shown. (Figure courtesy of \^Angela Santos)}
\label{sun}       
\end{figure}

While this chapter is dedicated to the theory of stellar pulsations, it is interesting and
motivating to start by inspecting one of the main observational
results in this context, namely, the oscillation power density
spectrum of the Sun. This is shown in the upper panel of 
Fig.~\ref{sun}. The first aspect that catches the eye is that the
oscillation spectrum  is composed of a number of discrete frequencies,
whose power is modulated over frequency, showing a close to Gaussian
shape. This is typical of oscillation spectra of solar-like
pulsators in which modes are intrinsically stable (meaning that small
perturbations are damped) and continuously excited stochastically by convection. Other stellar
pulsators, in which oscillations are intrinsically unstable, with small
perturbations growing due to some sort of coherent excitation
mechanism, will still show oscillation
spectra composed of discrete frequencies, but often a less regular
pattern as a consequence of not all
possible frequencies being excited or observed. An important observable for
solar-like pulsators is the frequency of maximum power, $\nu_{\rm max}$,
shown in Fig.~\ref{sun}. There are different approaches to derive it \citep[and references therein]{verner11}, that usually involve
considering a heavily-smoothed version of the oscillation power spectrum.

The lower panel of Fig.~\ref{sun} shows a close-up of the regular
peak structure seen in the upper panel. Here each mode is identified
by a positive integer, the mode degree, which will be discussed in
detail in Sect.~\ref{solutions}. Two main separations are identified in
the figure, namely, the {large separation}, $\Delta\nu$, between
consecutive modes of the same degree and the {small separation},
$\delta\nu$, between
modes of similar frequency and degree differing by two.

The large separation  has been shown to scale as
$\Delta\nu\propto\sqrt{\overline{\rho}}$ \citep{tassoul80},
where $\overline{\rho}$ is the mean density of the star. Moreover, the frequency of maximum power has been suggested to scale with the
surface gravity and effective temperature as $\nu_{\rm
  max}\propto~gT_{\rm
  eff}^{-1/2}$ \citep{Brown91,kjeldsen95}. Together, these scaling
relations provide two equations that can be used  for a first estimate
of the stellar mass and radius, once the effective temperature is known, namely,
\begin{eqnarray}
\frac{R}{{\rm R}_\odot}&\approx&\left(\frac{\nu_{\rm max}}{\nu_{\rm
      max,\odot}}\right)\left(\frac{\overline{\Delta\nu}}{\overline{\Delta\nu_\odot}}\right)^{-2}\left(\frac{T_{\rm
  eff}}{T_{\rm eff,\odot}}\right)^{1/2}\,, \nonumber  \\
\frac{M}{{\rm M}_\odot}&\approx&\left(\frac{\nu_{\rm max}}{\nu_{\rm max,\odot}}\right)^3 \left(\frac{\overline{\Delta\nu}}{\overline{\Delta\nu_\odot}}\right)^{-4}\left(\frac{T_{\rm
  eff}}{T_{\rm eff,\odot}}\right)^{3/2}\,, 
\label{scaling}
\end{eqnarray}
where the overbar stands for a suitable average taken over the different
pairs of modes and solar values are marked by the index `$\odot$'.

Finally, the small separation depends strongly on the sound speed in the
stellar core and is, thus, very sensitive to stellar age. These
quantities shall be discussed further in
Sect.~\ref{conclusions}. Before that, I will introduce the pulsation
equations in Sect.~\ref{problem}, and
discuss the corresponding solutions in Sects.~\ref{trapping}--\ref{numerical}.

\section{Equations for linear, adiabatic stellar pulsations}
\label{problem}

In this section I set the problem of linear, adiabatic stellar
pulsations. I start from the equations of hydrodynamics for an
inviscid fluid and then consider small, adiabatic perturbations about a
spherically symmetric, static equilibrium. Finally, I discuss the
functional form of the solutions on the sphere and the boundary conditions.


 \begin{figure}[t]
\sidecaption
\includegraphics[scale=.75]{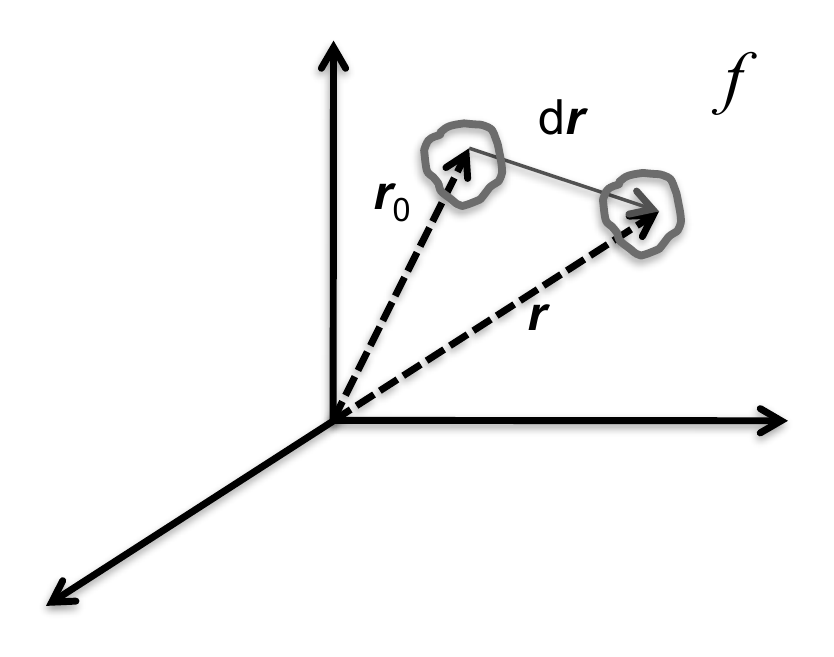}
%
%
\caption{Eulerian versus Lagrangian descriptions. In the Eulerian description
  the evolution of the property, $f$, of the gas is considered at
  fixed position, ${\bfr_0}$, by comparing $f({\bfr_0},t_0)$ with  $f({\bfr_0},t_1)$, while in
  the Lagrangian description the evolution is considered following the
  motion by comparing $f({\bfr_0},t_0)$ with  $f({\bfr},t_1)$. }
\label{fig1}       
\end{figure}

Let us assume that a gas can be treated as a continuum with thermodynamic
properties well defined at each position in space, $\bfr$. Let $f$ be a scalar
property of the gas. There are two ways of looking at the time evolution
of $f$: (i) at fixed position $\bfr_0$  and (ii) following the
motion (see Fig. \ref{fig1}). The
first corresponds to an Eulerian description and the second to a
Lagrangian description. Both perspectives are useful and commonly used
in the study of stellar pulsations. The two descriptions are related by
\begin{equation}
\frac{{\rm d}f}{{\rm d} t} = \frac{\partial f}{\partial
  t}+\nabla f\cdot\frac{\rm d \bfr}{\rm d t} \equiv\frac{\partial f}{\partial
  t}+\bfv\cdot\nabla f \, ,
\label{euler}
\end{equation}
where $\bfv$ is the velocity, d/d$t$ is the time derivative following the motion
(Lagrangian description) and $\partial /\partial t$ is the
time derivative at fixed position (Eulerian description).
Likewise, for a vector quantity, ${\textbf{\itshape F}}$, the two derivatives are related by
\begin{equation}
\frac{{\rm d}{\bff}}{{\rm d} t} = \frac{\partial{\textbf{\itshape F}}}{\partial
  t}+\left(\bfv\cdot\nabla\right)\bff\,.
\label{euler}
\end{equation}
%


\subsection{The conservation laws}
\label{equations}
The evolution of the properties of a fluid is described by a set of equations that
translate conservation laws. In what follows these equations are summarized under particular conditions that will be discussed below.
Conservation of mass, linear momentum, and
energy are  expressed, respectively,  by
\begin{eqnarray}
\frac{\rm d\rho}{{\rm d} t} &=& -\rho\nabla\cdot{\rm\bfv}\,,\nonumber\\
\rho \frac{\rm d\bfv}{{\rm d} t} & = & -\nabla p+\rho\bfg+{\bff_{\rm
    oth}}\,,\nonumber\\
\frac{{\rm d} q}{{\rm d} t} & = &\frac{{\rm d} E}{{\rm d} t}+p \frac{{\rm d} \left(1/\rho\right)}{{\rm d} t}\,,
\label{hydro}
\end{eqnarray}
where $\rho$ and $p$ are, respectively, the fluid density and pressure,
$\bfg$ is the acceleration of gravity, ${\bff_{\rm oth}}$ are
other body forces, expressed per unit volume, that may act on the fluid, besides gravity (e.g., the Lorentz force,
if a magnetic field is present), and $E$ and $q$ are, respectively,  the internal
energy and heat supplied to the system, both per unit mass. 

The first of these equations, known as the continuity equation,
expresses that
the rate of change of the mass within a given volume must equal, with opposite
sign, the mass crossing the surface that encloses that volume, per unit
time. The
second, the equation of motion, expresses that the change in linear
momentum of an element of fluid must equal the force acting on it by
its surroundings. It is written under the assumption that the fluid is
inviscid, which is a good approximation under stellar conditions.  The
third equation translates the first law of thermodynamics and it
states that the change in the internal energy of a system must equal
the heat supplied to the system minus the work done by the system on
its surroundings.  This equation can be written in different forms. A
useful one,  adopted below, is
\begin{equation}
\frac{{\rm d}{ q}}{{\rm d} t} =
\frac{1}{\rho\left(\Gamma_3-1\right)}\left(\frac{{\rm d} p}{{\rm d}
    t}-\frac{\Gamma_1 p}{\rho}\frac{{\rm d} \rho}{{\rm d}
    t}\right)\,,
\label{energy}
\end{equation}
where $\Gamma_1$ and $\Gamma_3$ are adiabatic exponents defined by the
adiabatic derivatives,
\begin{equation}
\Gamma_1=\left(\frac{\partial {\rm ln} p}{\partial {\rm ln}
    \rho}\right)_{\rm ad } \;\;,\hspace{1cm} \Gamma_3-1=\left(\frac{\partial {\rm ln} T}{\partial {\rm ln}
    \rho}\right)_{\rm ad } ,
\end{equation}
and $T$ is the temperature of the fluid. From Eq.~(\ref{energy}) one
can further define the adiabatic sound speed, $c$. Making the left-hand side equal to zero one finds:
\begin{equation}
c^2\equiv\frac{{\rm d} p}{{\rm d}\rho}=\frac{\Gamma_1 p}{\rho}\,.
\end{equation}

Finally, it should be noted that the thermodynamic variables $T$, $\rho$, and $p$ are not all independent, but
rather are related by the equation of state that can be expressed as
$\mathcal F \left(T, p,\rho\right)=0$, where $\mathcal F$ is a function
that depends on the conditions of the fluid. Since in this chapter only adiabatic oscillations
will be considered, the explicit
form of the equation of state will not be needed. However,
this equation will still be required if the reader is interested in
deriving the temperature fluctuations associated to the perturbations in the density and pressure.

\subsection{Perturbative analysis}
Consider an equilibrium state that is: (i) static, meaning that there
are no velocities and all derivatives at fixed position are null
($\partial /\partial t =0$) and (ii) spherically symmetric, implying,
e.g.,
that there is no rotation or magnetic fields. Then, in the equilibrium, one has:
\begin{equation}
\nabla p_0=\rho_0\bfg_0\equiv-\rho_0 g_0 {\rm{\hat{a}}_{\rm
        r}}\,,
\label{hydrostatic}
\end{equation}
where the index `0' is used to identify the equilibrium quantities and ${\rm{\hat{a}}_{\rm
        r}}$ is the unit vector in the radial direction pointing
    outwardly from the centre of the star,  making the scalar $g_0$ 
   a positive quantity. 

Now let us assume that the equilibrium is
perturbed under the following conditions: (i) the perturbations are
adiabatic and (ii) they are small, in the sense that non-linear terms
in the perturbations can be neglected.

The adiabatic condition implies the assumption that no heat is exchanged with the
element of fluid during the perturbation, a condition that  is very closely
satisfied almost everywhere in the star. This can be seen by comparing
the characteristic timescales of pulsations, typically found in the
range of minutes to a few days, with the timescale for radiation that, except very close to the
stellar surface, has characteristic values many orders of magnitude
larger than the pulsation period (e.g., exceeding a million years in
the Sun, when the Sun is considered as a whole).

Under the above conditions, let  $f$ be a
scalar property of the gas, and $f'$ and $\delta f$ be,
respectively, the Eulerian and Lagrangian  perturbations to it. Then
$f=f_0+f'$ and $\delta f=f '+ \boldsymbol{\xi}\cdot\nabla f_0$,
where  $\boldsymbol{\xi}$ is the displacement vector ($\equiv\bfr-\bfr_0$). Moreover, since the
perturbations are linear, the velocity of a given element of fluid is
\begin{equation}
\bfv \equiv\frac{{\rm d} {\boldsymbol{\xi}}}{{\rm d} t}\approx\frac{\partial{\boldsymbol{\xi}}}{\partial t}\,.
\label{vel}
\end{equation}

Perturbing the system of equations (\ref{hydro}) (with energy
conservation expressed as in  Eq.~\ref{energy}), using Eqs.~(\ref{hydrostatic}) and (\ref{vel}), and integrating in
time the equations of continuity and energy, one finds that linear adiabatic
perturbations about a static spherically symmetric equilibrium are described by the following set of equations:
\begin{eqnarray}
\rho ' &=&- \nabla\cdot\left(\rho_0{{\boldsymbol{\xi}} }\right)\,, \nonumber\\
\rho_0 \frac{\partial^2{\boldsymbol{\xi}}}{\partial t^2} &=& -\nabla
p'-\rho_0\nabla\phi '-\rho '\nabla\phi_0\,,\nonumber\\
\nabla^2\phi '& =& 4\pi G\rho '\,,\nonumber \\
p'+{\boldsymbol{\xi}}\cdot\nabla p_0 &=& \frac{\Gamma_{1,0}
  p_0}{\rho_0}\left(\rho'+{\boldsymbol{\xi}}\cdot\nabla\rho_0\right)\,.
\label{pert}
\end{eqnarray}
To reach the system of equations above, I have further defined the acceleration of gravity in
terms of the gravitational potential $\phi$, such that ${\bf
  g}=-\nabla\phi$ and, accordingly, considered in addition the
Poisson equation that relates the gravitational potential to the fluid
density. 

Taking the equilibrium quantities as known, one can identify four
variables in the system above (three scalars and one vector), namely
$\rho '$, $p'$, $\phi '$, and ${\boldsymbol{\xi}}$. These four equations
thus form a closed system that can be solved, with adequate boundary
conditions.

\subsection{Solutions on a sphere\label{solutions}}
Consider the spherical coordinate system $(r,\theta,\varphi )$ such
that the
variables $\rho ', p', \phi ',{\boldsymbol{\xi}}$ are expressed as functions of
$r,\theta,\varphi$ and $t$.

It can be shown by substitution (or derived by the technique of
separation of variables) that the system of equations (\ref{pert}) admits
solutions of the type
\begin{eqnarray}\label{sol}
f'\left(r,\theta,\varphi,t\right)&=&\Re\left\{{\textstyle
    f'\left(r\right)Y_l^m\left(\theta,\varphi\right){\rm  e}^{-i\omega
      t}}\right\}\,,  \\
\hspace{-0.2cm}{\boldsymbol{\xi}}\left(r,\theta,\varphi,t\right) &=&\Re\left\{\textstyle{\hspace{-0.05cm}\left[\xi_r\left(r\right)
    Y_l^m\left(\theta,\varphi\right){\rm{\hat{a}}_{\rm
        r}}+\xi_h\left(r\right){\left(\frac{\partial Y_l^m}{\partial\theta}{\rm{\hat{a}}_{\rm \theta}}+\frac{1}{\sin\theta} \frac{\partial Y_l^m}{\partial\varphi}{\rm{\hat{a}}_{\rm \varphi}}\right)}\right]{\rm  e}^{-i\omega t}\hspace{-0.1cm}}\right\}\,,\nonumber
\end{eqnarray}
where $f'$ stands for any of the scalar perturbations, $\xi_r$ and
$\xi_h$ are, respectively, the depth-dependent amplitudes of the radial and horizontal
components of the displacement and $\rm{\hat{a}_i}$ are the components
of the unit vectors of the spherical coordinate system.

The time dependence of the solution is associated to the angular oscillation frequency
$\omega$. The sign in the exponential is arbitrary. Here it is chosen to
be negative to guarantee that in cases when $\omega$ is complex the
growth rate (i.e., the imaginary part of the frequency) is positive when
the perturbation grows.  The possible values of $\omega$ are determined by imposing  the
boundary conditions that shall be discussed later. In practice, since the equations were derived under the
assumption that the perturbations are adiabatic, so far as the boundary conditions are fully
reflective (i.e., no energy is lost through the boundary), $\omega$ is
real. That will be the only case discussed in this chapter.

\begin{figure}[t]
\includegraphics[scale=.35]{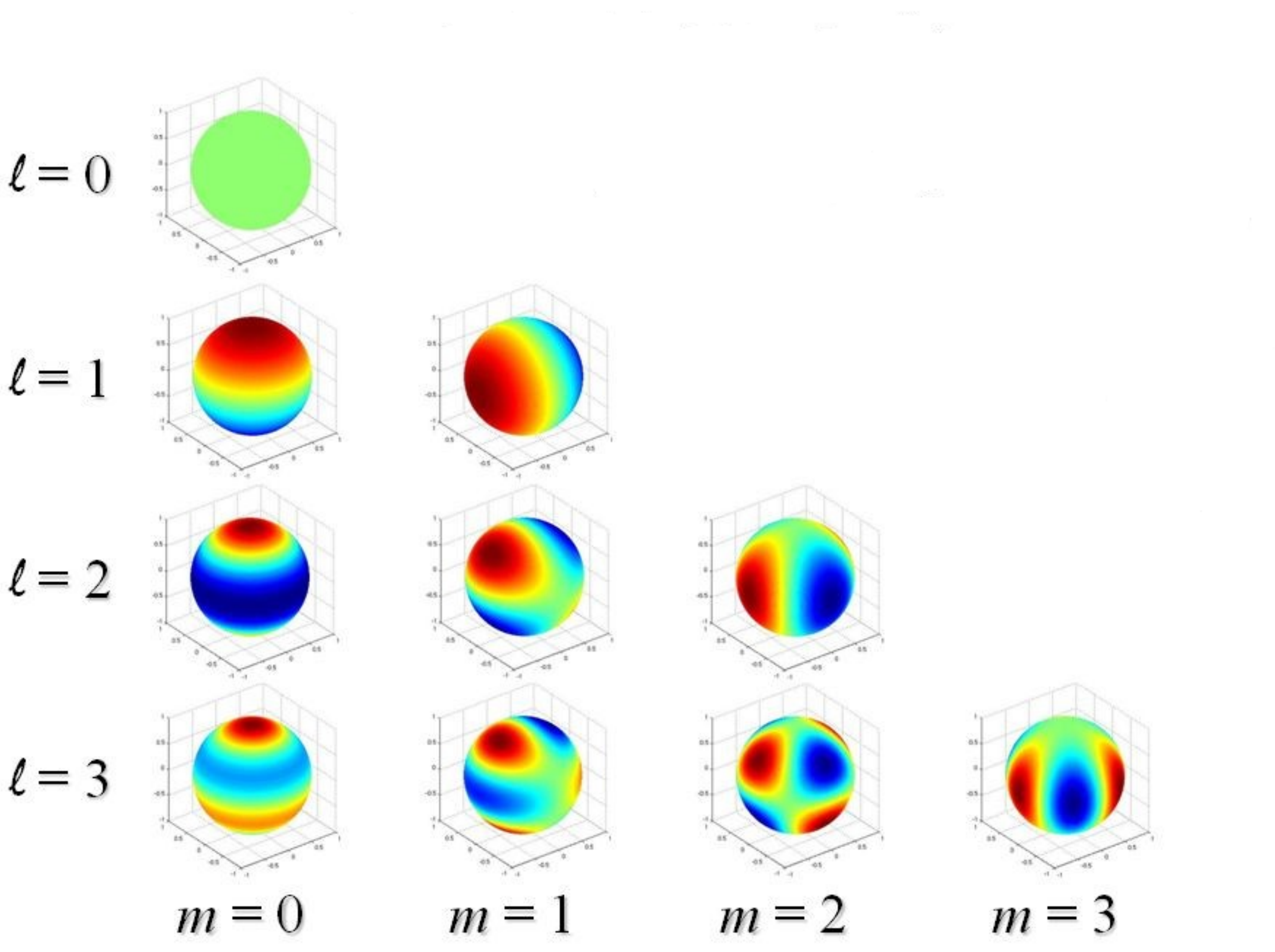}
%
%
\caption{Examples of spherical harmonic functions, $Y_l^m$, for mode
  degrees in the range $l=0$--$3$. For each mode degree, all possible non-negative
  values of $m$ are shown. Red and blue
  show perturbations of opposite sign.}
\label{Ylm}       
\end{figure}

 The angular dependence of the solutions is given by the
spherical harmonic functions $Y_l^m$, characterized by the angular
degree $l$ (a non-negative integer), and the azimuthal order $m$,
an integer that takes values between $-l$
and $l$. The angular degree
defines the number of surface nodes and the absolute value of the
azimuthal order defines the subset of those that cross the
equator. This means that $|m|$ defines the orientation of that solution
on the sphere, something that will be relevant for the discussion below. An
example of low-degree spherical harmonic functions with identified values of
$l$ and $|m|$ is shown in
Fig.~\ref{Ylm}. Since $Y_l^m\propto P_l^m(\cos\theta){\rm
  e}^{im\varphi}$, where $P_l^m$ are the associated Legendre
functions, the sign of $m$ defines whether the associated
solution is travelling eastwardly or westwardly
in the chosen reference
frame. Given the negative sign adopted for the time-dependent part of
the solution the perturbations are found to vary as ${\rm e}^{i(m\varphi-\omega
  t)}$. This means that in this case a positive $m$ corresponds to a
solution travelling eastwardly. I note, however, that not all
literature adopts the same definition, since sometimes the opposite
sign is chosen for the exponent in the time-dependent exponential. Finally, from the properties of
the spherical harmonics, one has that
\begin{equation}
\nabla_h^2 Y_l^m = -\frac{l\left(l+1\right)}{r^2}Y_l^m\equiv -\kappa_h^2 Y_l^m\,,
\label{kh}
\end{equation}
where $\kappa_h$ has been identified as the horizontal wavenumber of the
perturbation when the latter is interpreted, locally, as a plane wave.

\begin{figure}[t]
\hspace{0.5cm}\includegraphics[scale=0.37]{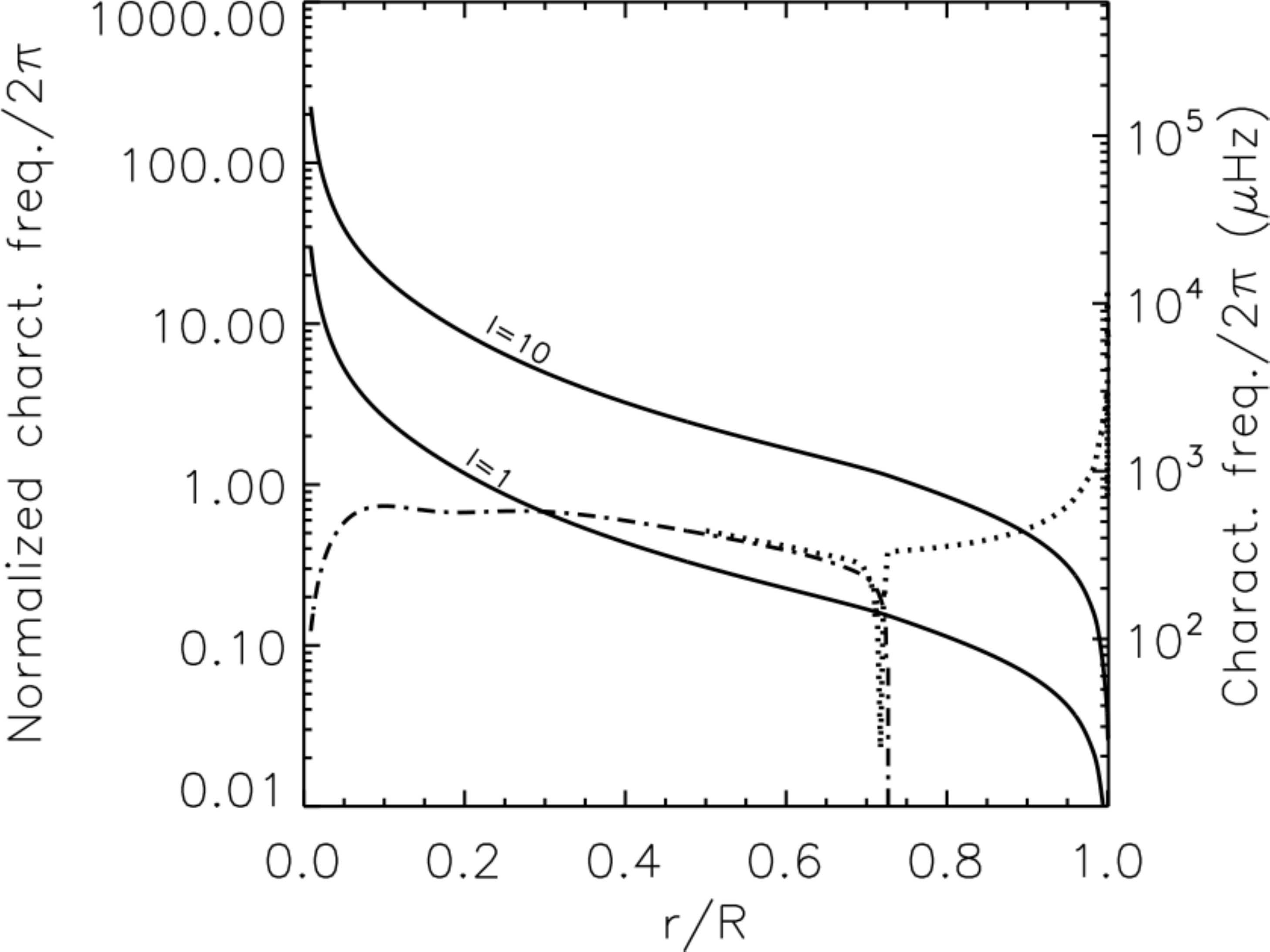} \hspace{0.5cm}

%
%
\caption{Lamb frequency, $S_{l}$ (continuous lines), for $l=1$ and
  $l=10$, buoyancy frequency, $N_0$ (dashed-dotted line), and
  critical frequency, $\omega_c$ (dotted line, displayed in the outer
  50$\%$ of the stellar radius only), for a model of the Sun, all divided by 2$\pi$. The
  left vertical axis shows dimensionless values of these characteristic
  frequencies obtained by multiplying them by $t_{\rm dyn}$. The
  right vertical axis indicates the true physical values. 
   }
\label{freqS}       
\end{figure}

\begin{figure}[t]
\hspace{0.5cm}\includegraphics[scale=0.37]{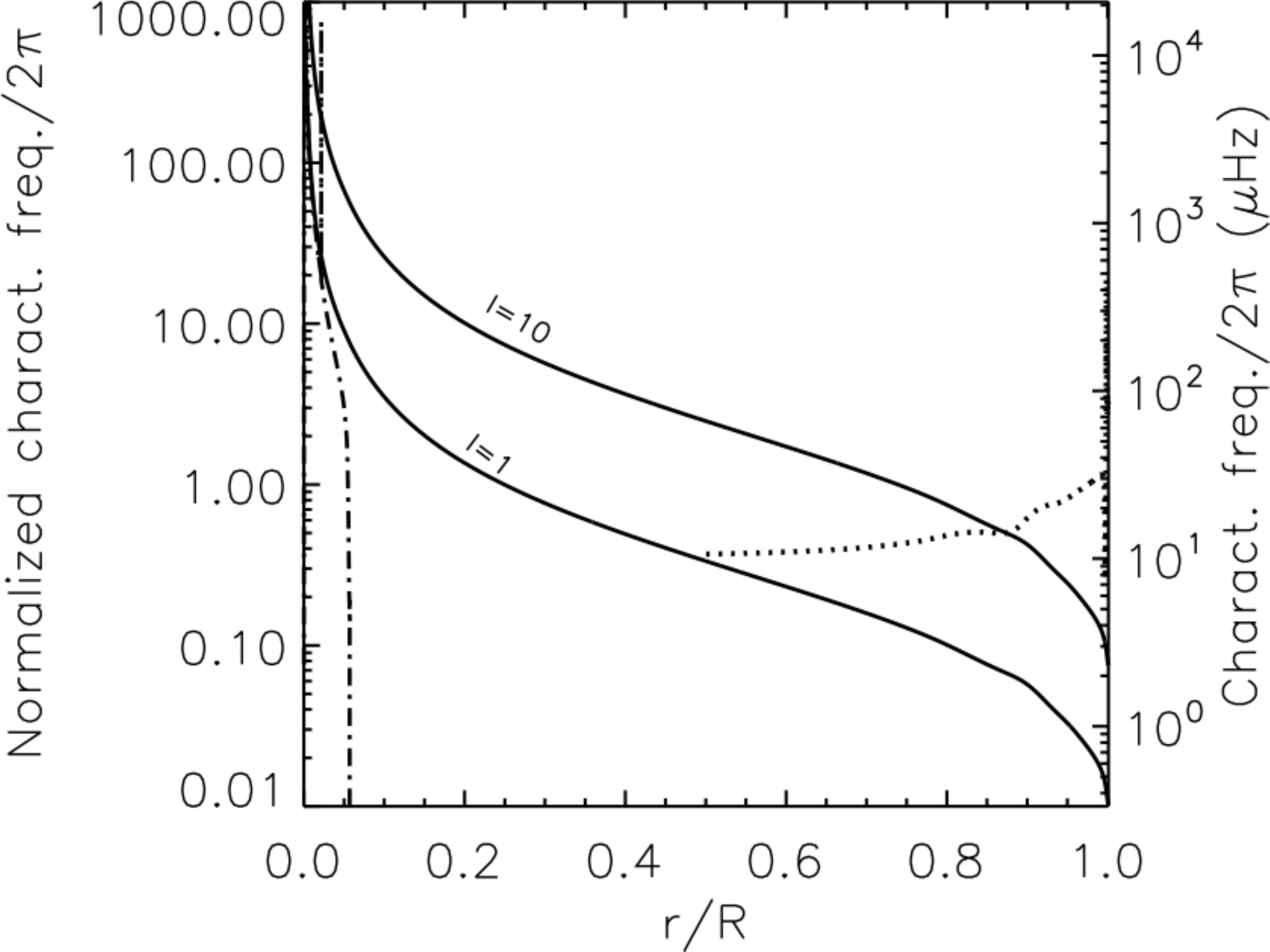}

%
%
\caption{Same as Fig.~\ref{freqS} but for a model of a star in the red-giant branch.
   }
\label{freqRG}       
\end{figure}

The last part of the solutions in Eq.~(\ref{sol}) are the radial-dependent
amplitudes. For simplicity, radial-dependent parts of scalar perturbations have been
named with the same symbol as the full
solutions. Using the full solutions separated as in
Eq.~(\ref{sol}) in the system of equations (\ref{pert}), one can derive a
set of equations governing these amplitude functions.  After eliminating the horizontal component of the displacement by
combining the continuity equation and the horizontal
divergence of the perturbed momentum equation, and eliminating the Eulerian
perturbation to the density through the adiabatic relation, one finds
that the radial-dependent amplitudes $p'(r)$, $\phi '(r)$, and $\xi_r(r)$, obey the following system of equations:
\begin{eqnarray}
\frac{1}{r^2}\frac{{\rm d}}{{\rm
    d}r}\left(r^2\xi_r\right)-\frac{g_0}{c_0^2}\xi_r
-\left(\frac{S_l^2}{\omega^2}-1\right)\frac{1}{c_0^2\rho_0}p' &=&
\frac{l\left(l+1\right)}{r^2\omega^2}\phi ' \,,\nonumber \\
\frac{{\rm d}p'}{{\rm
    d}r}+\frac{g_0}{c_0^2}p'-\rho_0\left(\omega^2-N_0^2\right)\xi_r&=&-\rho_0\frac{{\rm
  d}\phi '}{{\rm d}r} \,,\nonumber \\
\frac{1}{r^2}\frac{{\rm d}}{{\rm d}r}\left(r^2\frac{{\rm
      d}\phi '}{{\rm d}r}\right)-\frac{l\left(l+1\right)}{r^2}\phi
' &=&4\pi G\left(\frac{p'}{c_0^2}+\frac{\rho_0 N_0^2}{g_0}\xi_r\right)\,,
\label{amp}
\end{eqnarray}
where two characteristic frequencies have been defined: the Lamb
frequency, $S_l$; and the buoyancy (or Brunt--V\"ais\"al\"a) frequency,
$N_0$. The squares of these quantities are given, respectively, by 
\begin{equation}
S_l^2=\frac{l\left(l+1\right)c_0^2}{r^2} \;\;, \hspace{1cm}
N_0^2=g_0\left[\frac{1}{\Gamma_{1,0}}\frac{{\rm d}\ln p_0}{{\rm d}r}-\frac{{\rm d}\ln\rho_0}{{\rm d}r}\right]\,.
\end{equation}
 Examples of the Lamb frequency and buoyancy frequency are shown
 in Figs.~\ref{freqS} and \ref{freqRG} for a model of the Sun and a
 model of a
 star in the red-giant branch, respectively. The buoyancy frequency is
seen only where $N^2>0$,  which marks stellar layers that are stable to convection.

A significant difference is found in the
 characteristic values of the Lamb frequency in the two stars, as seen
by comparing the scales on the right-hand side vertical axes in
Figs.~\ref{freqS} and \ref{freqRG}. This
 is because this frequency scales approximately with the inverse of the dynamical timescale of
 the star, $t_{\rm dyn}=(R^3/GM)^{1/2}$, hence depending significantly
 on stellar radii. The vertical scale on the left-hand side of
 each figure shows the characteristic frequencies in units of $1/t_{\rm
 dyn}$, illustrating the similarity of the Lamb frequency in the two stars once the scaling is accounted
for. The second aspect that calls attention in these figures is
that the buoyancy frequency in the more evolved star shows a much more
significant contrast, increasing significantly towards the stellar
centre. This is a consequence of the increasing density gradient in
the innermost layers as the star evolves and the core contracts. These structural
differences between main-sequence and red-giant stars have
significant impact on the properties of their oscillations, as will be
discussed in Sect.~\ref{numerical}. 

There are a number of points that should be stressed in relation to
the system of equations (\ref{amp}):
\begin{itemize}
\item First, the system only contains total
derivatives, reminding us that the variables are the depth-dependent
parts of the solutions only. In fact, the system of equations (\ref{amp}) forms a system of linear, total
differential equations of 4th order for four unknown functions, namely, the depth-dependent
amplitude functions, $\xi_r, p ', \phi '$, and  $\rm{d}\phi
'/\rm{d}r$. We have, thus, at this point, reduced the original 3-dimensional problem
into a 1-dimensional problem.

\item The second point is that in the case of spherically symmetric
  perturbations ($l=0$), the perturbed Poisson equation can be
  integrated, reducing the system to 2nd order for the variables $p '$ and $\xi_r$. An important consequence, that will not be explored
  here, is that it is then possible to combine the two first-order differential equations to obtain a single
  second-order differential equation for the displacement, that can then be
  cast in the form of a standard wave equation. Moreover,
  \citet{takata05,takata16a} has shown that in the case of dipolar modes ($l=1$),
 momentum conservation can be used to derive an integral
  that  again reduces the system
  to 2nd order. In the same work, the author has also argued that
  integrals such as those found for the case of radial and dipolar modes
  do not exist for any other mode degree. 

\item For other mode degrees, reduction of the system of equations (\ref{amp}) to a 2nd
  order system can be achieved by performing the {\it Cowling approximation}, which consists in neglecting
  the perturbation to the gravitational potential. This approximation
  is adequate for perturbations that vary on relatively short scales
  (much smaller than the radius of the star). Since  the integral
  solution of the Poisson equation relates $\phi  '$ to the integral of
  $\rho '$, the cancellation effect that results from the integration
  when $\rho '$ varies on short scales, leads to $\phi
  '$ being small in that case. While this reduction of the order of the
  system requires an approximation, it has proven to be extremely useful
  for the asymptotic analysis of the equations and, in that way, a better understanding
  of the physical picture involved. I will get back to this in
  Sect.~\ref{trapping}.

\item The fourth, and last point, that should be made is that while the
  coefficients in the system of equations (\ref{amp})
depend on the mode degree, $l$, they are independent of the mode
azimuthal order, $m$. This is a consequence of our assumption that the
equilibrium state is spherically symmetric. In fact, under that
assumption, there is no preferential direction in the star and, so far
as the boundary conditions (to be discussed in Sect.~\ref{bc}) are also
independent of $m$, the solutions must be independent of the reference about which the
spherical harmonic functions are defined. Since $m$ defines the
orientation of the spherical harmonic on the sphere, the equations and,
hence, the solutions, must not depend on $m$. This means that under
this assumption the solutions will be degenerate in the azimuthal
order. That degeneracy, however, is broken (partially broken), in the presence of agents
that break (partially break) the spherical symmetry, such as, e.g.,
rotation or magnetic fields.

\end{itemize}

\subsection{Boundary conditions\label{bc}}
Of the four boundary conditions required to solve the problem set by the system of equations (\ref{amp}),
two will be defined at the stellar centre ($r=0$) and two at the stellar
surface ($r=R$). \\

\subsubsection{ At the stellar centre}

The boundary conditions at the stellar centre are
derived by imposing that the solutions are regular (do not diverge) there. By expanding
the equations near $r=0$ one finds that the regular solutions require
$p '\sim \mathcal{O} (r^l)$, $\phi  '\sim \mathcal{O} (r^l)$, and
$\xi_r\sim \mathcal{O} (r^\alpha)$, where $\alpha=1$ for $l=0$ and
$\alpha=l-1$ for $l>0$. This means that, as $r \rightarrow 0$, 
\begin{equation}
\frac{\rm{d}\phi '}{\rm{d}r}-\frac{l}{r}\phi '\rightarrow 0 \;\;,
\hspace{0.5cm}\frac{\rm{d}p '}{\rm{d}r}-\frac{l}{r}p '\rightarrow 0 \;\;,
\hspace{0.5cm}\frac{\rm{d}\xi_r}{\rm{d}r}-\frac{\alpha}{r}\xi_r \rightarrow 0\,.
\label{bc0}
\end{equation}
The three conditions above are not all independent. In
fact, the condition on the displacement
can be derived from the other two by first noting
that the gradient of the thermodynamic variables in the equilibrium
structure must be zero at the centre of the star and, then, applying that knowledge to
the perturbed equations. Moreover, in that process one also finds that
for non-radial modes $\xi_r=l\xi_h$, at $r=0$.  The regularity of the solutions
thus provides us with two independent boundary conditions to apply at
$r=0$. One interesting point to notice is that the displacement at the
centre of the star is non-zero only for dipolar modes ($l=1$). In all
other cases the centre of the star does not move. That such is the
case can also be seen from symmetry arguments, as only for $l=1$ modes
one would recover the same non-zero displacement vector at $r=0$ independently from
where the centre is approached. \\

\subsubsection{At the stellar surface\label{bcup}}

As the density vanishes outside the star, at the surface the perturbation to the
gravitational potential must match continuously onto
the physically-meaningful solution of the Poisson equation for a
vacuum field (vanishing at
infinity). That implies that $\phi  '\sim \mathcal{O} (r^{-l-1})$ and
thus, at $r=R$,
\begin{equation}
\frac{\rm{d}\phi '}{\rm{d}r}=-\frac{l+1}{r}\phi '\,.
\label{bc1}
\end{equation}

The second boundary condition to be applied at the surface depends on
how one treats the atmosphere of the star. If one assumes a free
boundary, then one must consider that the pressure at the boundary is constant
and, hence, that the Lagrangian pressure perturbation there is
zero. In that case,  a second boundary condition at $r=R$ is found in
the form:
\begin{equation}
p'+\frac{{\rm d}p_0}{{\rm d}r}\xi_r=0\,.
\label{bc2}
\end{equation}
This condition is reasonable for low-frequency waves, as will be seen later. However, as the frequency of the waves increases, the details of the atmosphere become
more important for the solution. It is therefore common, when solving
the pulsation equations, to adopt a more adequate boundary condition,
such as that derived from the matching of the radial displacement
solution onto the physically meaningful  analytical solution derived
for an isothermal atmosphere. The analytical solution can be derived
assuming a plane-parallel isothermal equilibrium composed of an ideal
gas and with constant adiabatic exponent and mean molecular
weight. Under these assumptions, the sound speed is constant and so
are the density and pressure scale heights, i.e., the characteristic
lengths associated with the variations of density and pressure,
given respectively by $H=-{\rm  d} r/{\rm d}\log\rho_0$ and $H_p=-{\rm
  d} r/{\rm d}\log p_0$. Moreover, in this case, the equilibrium pressure and
density decrease exponentially with height in the atmosphere
$\rho, p \propto\exp [\small{(R-r)/H)}]$. Then, considering the system of equations (\ref{amp}) under the Cowling
approximation, the displacement is found to have the form $\xi_r\propto\exp(\kappa r)$, where,
for all cases that will be of interest to us\footnote{Here the term that would dominate in the case of
  atmospheric gravity waves is being neglected, as those
  will not be discussed in these lectures.},
\begin{equation}
\kappa=\frac{1}{2H}\left[1\pm\left(1-\frac{4\omega^2H^2}{c_0^2}\right)^{1/2}\right]\equiv\frac{1}{2H}\left[1\pm\left(1-\frac{\omega^2}{\omega_{\rm
      c}^2}\right)^{1/2}\right]\,.
\label{kbc}
\end{equation}
Combining the perturbed equation of mass conservation and the
adiabatic condition (first and fourth equations in the system of
equations (\ref{pert})) with the
solution for $\xi_r$, and noting that in the isothermal atmosphere
$H=H_p$, one finds an alternative boundary condition to be applied
at $r=R$ (replacing Eq.~\ref{bc2}), namely,
\begin{equation}
p '\approx -\left(\Gamma_{1,0}p_0\kappa+\frac{{\rm d}p_0}{{\rm
      d}r}\right)\xi_r=\frac{p_0}{H}\left(1-\Gamma_{1,0}\kappa H\right)\xi_r\,,
\label{bc3}
\end{equation}
where $\nabla\cdot{\boldsymbol\xi}$ has been approximated by ${\rm
  d}\xi_r/{\rm d}r$, implicitly assuming that the perturbation varies much
more rapidly in the radial than in the horizontal direction. 

In Eq.~(\ref{kbc}) the critical frequency, $\omega_c$, has been introduced,
which in the isothermal atmosphere is constant and equal to $c_0/\small{(2H)}$. It is important to note that when $\omega
< \omega_c$, $\kappa$ is real and the physically meaningful 
solution corresponds to choosing the negative sign, ensuring that
the energy density, $\propto\rho\xi_r^2$, decreases outwardly. When
$\omega > \omega_c$, $\kappa$ is complex and assuming no waves are
being sent into the atmosphere from outside, the imaginary part must
be chosen to guarantee that the wave travels outwardly.  Considering
the time-dependent part of the solution, one finds that the radial
component of the displacement goes as $\exp[i(\pm\kappa_i r-\omega
t)]$, where $\kappa_i=(\omega^2/\omega_c^2-1)^{1/2}$. The outwardly travelling solution is thus obtained by taking the
positive sign in Eq.~(\ref{kbc}). 

Under this boundary condition, waves with $\omega > \omega_c$ will
simply propagate away, loosing their energy through the
boundary\footnote{In a real stellar atmosphere there can be partial reflection of the wave
  energy even when $\omega > \omega_c$.  Accounting for that would
  require modifying the atmospheric model and, thus, the outer  boundary condition
  accordingly.}.  Here I am interested only in waves that are fully
  trapped inside the star, loosing no energy through the boundary,
  hence I will consider only the case when $\omega < \omega_c$.
  In the particular case when $\omega \ll \omega_c$ the expression
  for $\kappa$ can be expanded and the boundary condition (Eq.~\ref{bc3})
approximated by
\begin{equation}
p'=\frac{p_0}{H}\left(1-\Gamma_1\frac{\omega^2}{4\omega_{\rm c}^2}\right)\xi_r\,.
\label{bc3_a}
\end{equation}
Finally, I note that as the frequency decreases, the
second term inside the brackets on the right-hand side of Eq.~(\ref{bc3_a}) gets smaller
and this boundary condition approaches the one defined in
Eq.~(\ref{bc2}), justifying the adequacy of the latter in the case of
sufficiently low frequencies.

\subsection{Eigenvalues}

The system of equations (\ref{amp}) and associated boundary conditions
constitute an eigenvalue problem that needs to be solved numerically.
The system admits non-trivial solutions only for discrete
values of the eigenvalues $\omega$. The discrete solutions can be associated to
an integer number $n$, denominated by radial order. 
Once the depth-dependent amplitudes of the perturbations are computed, the full
solutions can be derived from Eq.~(\ref{sol}). 

In summary, the eigenvalues, $\omega=\omega (n,l,m)$, of the
3-dimensional problem set by the pulsation equations (\ref{pert}) (after separation
of time) are characterized by three quantum numbers, $n$, $l$, and $m$,
where the absolute value of the first, $|n|$, is related to the
number of nodes of the perturbation along the radial direction, while
the other two, introduced in Sect.~\ref{solutions}, are
related to the angular dependence of the solutions, in particular to
the horizontal scale of the perturbation and its orientation on the
stellar surface. The discrete nature of the eigenvalues is clearly
seen in the power density spectrum of the Sun shown in
Fig.~\ref{sun}.

As discussed before, in the absence of physical agents that break the
spherical symmetry of the problem, the solutions must be degenerate in
$m$. Hence, in that case one has  $\omega=\omega (n,l)$ and  any
linear combination of the $2l+1$ independent solutions associated with the
spherical harmonic of
degree $l$,  and different $m$ values, is still an
eigensolution for that eigenvalue.

A discussion of the full solutions obtained from numerical integration of the pulsation
equations will be presented in Sect.~\ref{numerical}. First, however,  it is useful
to analyse the second-order equation that is derived from the system of
equations~(\ref{amp}) under particular approximations. That will be
discussed in the next section.

\section{Trapping of the oscillations\label{trapping}}

The full solutions of the linear, adiabatic pulsation equations must be
computed numerically. Nevertheless, under the Cowling
approximation, valid for large absolute values of the radial order,
$|n|$, or for large degree $l$, the system of equations (\ref{amp}) reduces to 2nd order on the variables $p
'$ and $\xi_r$, namely, 
\begin{eqnarray}
\frac{1}{r^2}\frac{{\rm d}}{{\rm
    d}r}\left(r^2\xi_r\right)-\frac{g_0}{c_0^2}\xi_r
-\left(\frac{S_l^2}{\omega^2}-1\right)\frac{1}{c_0^2\rho_0}p' &=&0\,,\nonumber \\
\frac{{\rm d}p'}{{\rm
    d}r}+\frac{g_0}{c_0^2}p'-\rho_0\left(\omega^2-N_0^2\right)\xi_r
&=& 0\,.
\label{amp_cow}
\end{eqnarray}
These equations can be combined to find a second-order wave equation
for a single variable. To do so, I follow the work of Deubner and Gough
\citep{deubnerandgough84}, which, in addition to the Cowling approximation, assumes that locally the
oscillations can be treated as in a plane-parallel layer under
constant gravity, hence neglecting\footnote{For a more general case in which these assumptions are not
made, see \citet{gough93}.} the derivatives of $r$ and $g_0$.
Let us introduce a new variable,
\begin{equation}
\Psi=\rho_0^{1/2}c_0^2\nabla\cdot{\boldsymbol\xi}\,,
\end{equation}
which through the adiabatic condition and the continuity equation can
be seen to be  directly related to the Lagrangian pressure perturbation ($\delta
p=-\rho_0^{1/2}\Psi$). The second-order
system (\ref{amp_cow}), under the approximations mentioned above, can be
manipulated to derive a wave equation for $\Psi$, 
\begin{equation}
\frac{{\rm d}^2\Psi}{{\rm d}r^2}+\kappa_r^2\Psi = 0\,,
\label{wave}
\end{equation}
where $\kappa_r$ is the local radial wavenumber given by
\begin{equation}
\kappa_r^2= \frac{1}{c_0^2}\left[S_l^2\left(\frac{N_0^2}{\omega^2}-1\right)+\omega^2-\omega_c^2\right]\,,
\label{wavenumber}
\end{equation}
and the critical frequency is now given by
\begin{equation}
\omega_c^2=\frac{c_0^2}{4H^2}\left(1-2\frac{{\rm d}H}{{\rm d}r}\right)\,.
\label{wc}
\end{equation}
Notably, $k_r$ depends critically on the three characteristic
frequencies introduced before, namely, $S_l$, $N_0$ and
$\omega_c$. Examples of $\omega_c$ for a solar model and a red-giant
model are shown in
Figs.~\ref{freqS} and \ref{freqRG}, respectively. It is small in the stellar interior, where the
density varies on large scales, but it becomes large near the
surface, where structural variations take place on a much shorter
scale.

\subsection{Mode propagation cavities}
If $\kappa_r^2$
were constant and positive, the solution would be oscillatory
everywhere. However, in a star $\kappa_r^2$ is a
function of $r$,  and one may generally expect it to be positive in some
region(s) and negative in others. Where it is positive, the solution
is locally wave-like (oscillatory in $r$)  while where it is negative it is locally
exponential. Because $\kappa_r^2$ depends on $\omega$, the regions
where wave-like solutions are found will depend on the mode under
consideration.  When there is only one region of the star where
$\kappa_r^2 >0$, the mode is said to be trapped there and that
region is often called the mode propagation cavity. Away from that
cavity, where the energy of the mode decreases exponentially, the mode is said to
be evanescent. When $\kappa_r^2$ is positive in more that one region of the
star, separated by regions where it is negative, the mode propagates
in more than one cavity. However, often most of its energy 
is concentrated in one of these cavities whose structure, in turn, is the
most determinant for the properties of the mode. 

The radii at which $\kappa_r^2 =0$ are called the turning points and
define the edges of the propagation cavities. Setting the left-hand
side of Eq.~(\ref{wavenumber}) to zero, one finds a second-order
algebraic equation for $\omega^2$, whose roots,
$\omega_{l,\pm}^2$, are given by 
\begin{equation}
\omega_{l,\pm}^2=\frac{1}{2}\left(S_l^2+\omega_c^2\right)\pm\frac{1}{2}\sqrt{\left(S_l^2+\omega_c^2\right)^2-4S_l^2N_0^2}\,,
\label{omegapm}
\end{equation}
where the index $l$ was used to recall that the roots depend on the
mode degree through their dependence on the Lamb frequency.
Rewriting Eq.~(\ref{wavenumber}) as
\begin{equation}
\kappa_r^2= \frac{1}{c_0^2}\left[\omega^2-\omega_{l,+}^2\right]\left[\omega^2-\omega_{l,-}^2\right]\,,
\label{wavenumber_2}
\end{equation}
one sees that modes propagate if
\begin{equation}
\omega>\omega_{l,+} \hspace{0.5cm} {\rm or} \hspace{0.5cm}  \omega<\omega_{l,-}\,,
\end{equation}
 and modes are evanescent if
\begin{equation}
 \omega_{l,-}<\omega<\omega_{l,+}\,.
\end{equation}

\begin{figure}[t]
\includegraphics[scale=0.29]{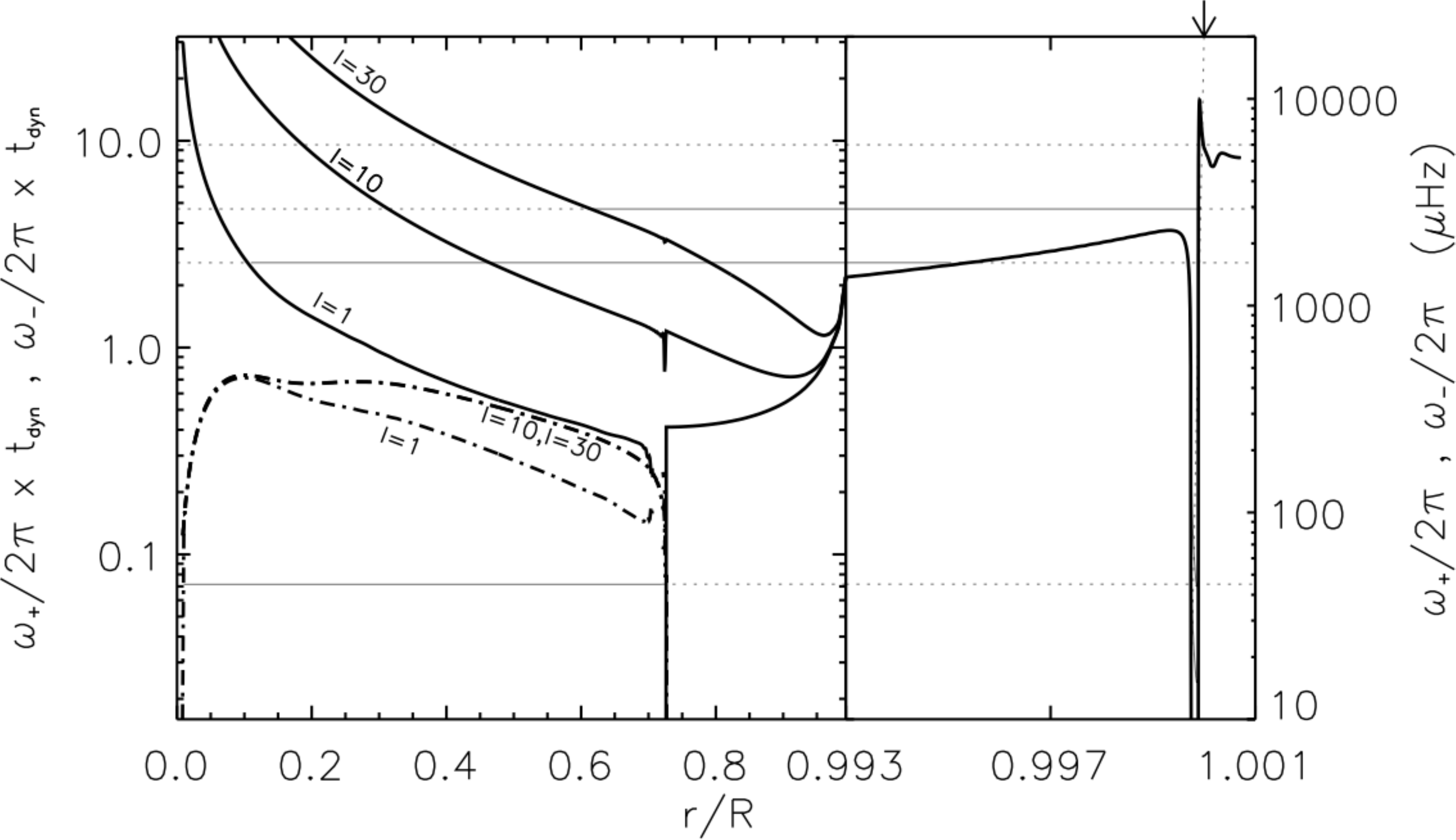}
%
%
\caption{Propagation diagram for model S of the
  Sun \citep{jcd96}. The frequencies $\omega_+$ (continuous thick lines)
  and $\omega_-$ (shaded-dotted thick lines) are shown for three
  different values of mode degree, $l$. Note that in the outer layers (right-hand
   section of the plot), the horizontal axis varies much slower than in
  the inner layers (left-hand section of the plot).  The
  arrow at the top marks $r/R=1$. The horizontal lines mark four
  different frequencies and are continuous where the mode is
  trapped. At the lowest end, we have a characteristic g-mode frequency
  for this model, with $\nu\sim 45\mu$Hz. At the highest end we have a
  frequency  $\nu\sim 6000\mu$Hz that is too high to be trapped
  inside the star. In between, we have two p-modes, one with
  $l=1, \nu\approx 1613\mu$Hz, and one with $l=30, \nu\approx
  2936\mu$Hz. The left-hand side vertical axis shows the range of
  values taken by these frequencies when they are scaled according to
  their dependence on the dynamical timescale, while the right-hand side axis
  shows their true physical values. }
\label{propg}       
\end{figure}

Figure \ref{propg} shows the frequencies $\omega_{l,\pm}$ for a model
of the Sun in what is usually called a propagation diagram.
Comparison with Fig.~\ref{freqS} shows
the resemblance between $N_0$ and $\omega_{l,-}$. The frequency
$\omega_{l,+}$, on the other hand, resembles $S_l$ in the deeper
layers of the stellar model, but it is clearly dominated by $\omega_c$ in the
outer layers.  Three modes, two at high frequency and one at low
frequency, are illustrated by the horizontal lines in
Fig.~\ref{propg}. For each of these modes, the propagation cavity 
corresponds to the section in which the horizontal line is
continuous. A fourth frequency is marked, with $\nu\sim 6000\:{\rm \mu Hz}$,
which is too high to correspond to a trapped mode. 

\subsection{Acoustic versus internal gravity waves}
Inspection of the propagation diagram shown in Fig.~\ref{propg} points
towards the existence of two families of solutions, one at lower
frequencies, where the mode cavity is essentially determined by the
buoyancy frequency, and one at higher frequencies, where the propagation cavity is
determined by the combination of the Lamb frequency, in the deeper
regions, and the critical frequency, near the surface. This can also
be seen by considering the lower and higher frequency limits of
$\kappa_r^2$ in the propagation region, as discussed below.

\subsubsection{Acoustic waves}
Let us consider first the higher frequency limit, namely, the case when
$\omega^2 \gg N_0^2$, in regions where $\kappa_r^2>0$.   Except near the surface, where
$\omega_c$ becomes large, Eq.~(\ref{wavenumber}) then gives
\begin{equation}
\kappa_r^2\approx\frac{\omega^2}{c_0^2}-\frac{l\left(l+1\right)}{r^2}\,.
\label{kracoustic}
\end{equation}
Recalling that the last term on the right-hand side is $\kappa_h^2$
(cf.~Eq.~\ref{kh}), one finds that
\begin{equation}
|\bfk|^2\equiv\kappa_r^2+\kappa_h^2\approx\frac{\omega^2}{c_0^2}\,.
\label{kacoustic}
\end{equation}
One thus has, for $\omega^2 \gg N_0^2$, a dispersion relation of the
type:
\begin{equation}
 \omega\approx c_0|{\bfk}|\,,
\label{dispacoustic}
\end{equation}
 which is characteristic of acoustic
waves. These waves are maintained by the gradient
of the pressure perturbation, i.e., the first term on the right-hand
side of the perturbed momentum equation. 
Note that in this case the frequency of the mode increases as
$\kappa$ increases. Taking the radial order for this family of
solutions as being positive integers, $n$, one thus finds that the
frequency of acoustic modes increases both with increasing radial
order, $n$, and with increasing degree, $l$.

Looking  back at Eq.~(\ref{kracoustic}),
one can further establish the lower turning point for the acoustic modes, that defines the
lower boundary of their propagation cavity. Setting $\kappa_r^2$ to
zero implies $\omega^2=S_l^2$, from which the lower
turning point is found to be defined by
\begin{equation}
r_{1,l}=\frac{\sqrt{l\left(l+1\right)}c_0{\scriptscriptstyle{\left(r_{1,l}\right)}}}{\omega}\,,
\label{ltp}
\end{equation}
where the subscript $l$ has been used to emphasize that the lower
turning point depends on the mode degree and it has been explicitly
indicated that the value of  $c_0$ is to be taken at the turning point. 
From Eq.~(\ref{ltp}), we see that the lower turning point of acoustic
modes depends strongly on the mode degree. As the mode degree
increases, the lower turning point gets closer to the surface,
implying that the propagation cavity of the mode becomes
shallower. That is also evident in Fig.~\ref{propg}, where one can see
that the depth at which a  horizontal
line in the high-frequency regime crosses lines of $\omega_{l,+}$ for different degrees, becomes
smaller as the degree increases.  In addition, we see from
Eq.~(\ref{ltp}), and also form inspection of Fig.~\ref{propg},  that
for a fixed mode degree, the lower turning point
decreases as the mode frequency increases. That means that higher
frequency acoustic modes propagate deeper, for fixed mode degree. 

The upper turning point of the acoustic modes is, in turn, determined
by comparing the oscillation frequency with the critical frequency,
which near the surface is much greater than $S_l$. There, one may approximate,
\begin{equation}
\kappa_r^2\approx\frac{\omega^2-\omega_c^2 }{c_0^2}\,.
\end{equation}
Thus, one finds that $\kappa_r =0$ in the outer layers if
\begin{equation}
\frac{c_0}{2H}\left[1-2\frac{{\rm d} H}{{\rm d}r}\right]^{1/2}\approx\omega\,,
\end{equation}
which provides an implicit condition for the upper turning point of
acoustic modes, $r_{2}$.  Note that unlike the case of the lower turning point, to this approximation the upper
turning point of acoustic modes is independent of the mode degree.

The upper turning point of acoustic modes in a solar-like model is
best seen in the right-hand section of 
Fig.~\ref{propg}. For the
lowest of the three frequencies in the high-frequency regime
($\nu\approx 1613\mu$Hz), the upper turning point is below the photosphere, while
for the second lowest ($\nu\approx
  2936\mu$Hz) it is at the photosphere. For that reason, the
former is significantly less sensitive to the
details of the outer layers than the latter. This is relevant because these layers are
particularly difficult to model. The oscillation frequencies derived
from models are therefore affected by the
incomplete modelling of the outer layers and, as a result, show
systematic differences when compared with the observations. This is
less so for the lower frequency acoustic modes, which may, in that case, serve
as an anchor with which to get a handle on the systematic errors
(assuming some kind of frequency dependence of these errors).
The other point to notice is that there is a frequency
above which no trapping is possible. This has been discussed in
Sect.~\ref{bcup}, where the critical
frequency for a plane-parallel, ideal, isothermal atmosphere has been introduced.  Independently of taking the
latter, or the more realistic critical frequency shown in Fig.~\ref{propg}, it is clear that bove $\sim 5.3$~mHz full trapping of
the modes no
longer occurs.
Since $\omega_c$ also scales with the inverse of the dynamical
timescale (as seen when comparing Figs.~\ref{freqS} and \ref{freqRG}),
the maximum expected observed frequency is strongly dependent
on the evolutionary state of the star. 

\subsubsection{Internal gravity waves}
Let us now turn our attention to the low frequency limit of
Eq.~(\ref{wavenumber}) in regions where $\kappa_r^2> 0$.  Let us consider
that $\omega^2\ll S_l^2$ throughout that propagation region 
and, in addition, that $\omega_c^2\ll
S_l^2$ there. In that case $k_r^2$ is given approximately by
\begin{equation}
\kappa_r^2\approx\frac{S_l^2}{c_0^2\omega^2}\left[N_0^2-\omega^2\right]\,.
\label{kgravity}
\end{equation}
Recalling that $\kappa_h^2=l(l+1)/r^2=S_l^2/c_0^2$, we then find the
dispersion relation
\begin{equation}
\omega^2\approx\frac{N_0^2}{1+{k_r^2}/{\kappa_h^2}}\,,
\label{dispgravity}
\end{equation}
which is characteristic of internal gravity waves. Internal gravity
waves are maintained by the gravity acting on the perturbation to the
density. If one considers a slow upwards displacement of an element of fluid whose pressure is
kept in equilibrium with the surrounding, buoyancy will respond to
restore the fluid towards the equilibrium position if its density is
larger than that of the surroundings.  That, in turn, can lead to the
oscillatory motion associated to gravity waves. Because an element of fluid
cannot  move strictly vertically, there is always an horizontal
component of the motion, which, in turn,
means that gravity waves can never be associated to spherically
symmetric perturbations, i.e., there are no gravity waves of
degree $l=0$. Moreover, since buoyancy will only oppose to the motion of
the element of fluid  where $N_0^2>0$, i.e., in convectively stable regions, the gravity waves will only
propagate where there is no convection. 

Looking back at the dispersion relation in Eq.~(\ref{dispgravity}), there are
additional points that should be noted. First,
the frequency of gravity waves is always smaller than $N_0$. This just confirms the
role of buoyancy in maintaining the dynamics of gravity waves. The
second point is that the frequency of gravity waves depends critically
on the shape of the perturbation. In fact, when the perturbation is
``needle-like'',  meaning that $K_r^2/K_h^2\ll 1$, the frequency is
higher, tending to $N_0$ as that ratio tends to zero. In the other
limit, for wide perturbations with  $K_r^2/K_h^2\gg 1$, the
oscillation frequency is smaller, tending to
zero as the ratio becomes increasingly higher. This can be understood if we recall that the amount of material
displaced horizontally is larger in the latter case than in the
former. This horizontal displacement, and the horizontal pressure
gradient that  it originates, increases the effective inertia of the
element of fluid on which the buoyancy force is acting. The
result is a smaller acceleration of the element of fluid and,
consequently, a smaller mode frequency.

It is worth emphasizing that the aspects discussed above are
in striking contrast with what was previously found for acoustic
waves. Indeed, for acoustic waves the frequency is found to depend essentially on the
characteristic scale of the perturbation, determined by the total wavenumber $|\boldsymbol\kappa|$, while for
gravity waves the frequency depends in addition, and very critically,  on the relation between the horizontal and
vertical scales. Moreover, considering modes of a given degree (hence
fixing the horizontal scale), we find that the frequency of gravity
waves decreases with increasing $\kappa_r$. Taking the radial orders
$n$ as negative integers, as is commonly done for gravity waves, we
see  that their frequencies decrease with increasing $|n|$, again in contrast with
what was found earlier for acoustic waves.

Finally, from Eq.~(\ref{kgravity}) we find that under the conditions
assumed, gravity waves propagate between the radii at which
$N_0=\omega$. The latter thus provides an implicit condition for the
lower and upper turning points of these modes which to this
approximation are independent of the mode degree, $l$. This is, again,
in contrast with the case of acoustic modes, for which the lower
turning point, and, hence, the extent of the propagation cavity, was
found to be strongly dependent on the mode degree. For the case of a star
like the Sun, we see from Fig.~\ref{propg} that the lower frequency gravity modes are
essentially trapped between the centre of the  star and the base of
the convective region. For the highest frequency gravity modes the
upper turning point gets smaller and the modes are trapped in deeper
layers. For main-sequence stars more massive than the Sun, the
innermost layers are convectively unstable and, thus, the cavity of gravity waves
is bounded on the inner side by the edge of the convectively unstable
core. On the other hand, since the convective envelope gets
shallower for more massive stars, gravity waves there can propagate almost
to the stellar surface. In more evolved stars, the trapping region
again depends on the existence, or not, of convection in the core, as
well as on the extent of the convective envelope, which can get very
deep, as happens, for instance, along the red-giant branch. Moreover,
the steep increase of the buoyancy frequency in the core of evolved
stars, such as that shown in Fig.~\ref{freqRG}, can lead to several cavities of
propagation for the same mode, in which case the analysis becomes more
complex than in the cases discussed above. I will get back to that case in the next section,
where  the full numerical results of the pulsation
equations shall be discussed. Table \ref{tab:1} provides a summary of the properties of acoustic
and gravity waves. 

\begin{table}[t]
\caption{Summary of the properties of acoustic and gravity waves discussed in
this chapter}
\label{tab:1}       
%
%
\begin{tabular}{ |p{5.5cm}|p{5.5cm}|  }
\hline
{\bf Acoustic waves} & {\bf Gravity waves} \\
\svhline
\fontsize{8}{5} \selectfont{Maintained by the gradient of pressure fluctuation} & \fontsize{8}{5}\selectfont {Maintained by gravity
acting  on density fluctuation}  \\
\hline
\fontsize{8}{5} \selectfont{Radial or non-radial} &
\fontsize{8}{5}\selectfont {Always non-radial}  \\
\hline
\fontsize{8}{5} \selectfont{Propagate in convectively stable or unstable regions} &
\fontsize{8}{5}\selectfont {Propagate
in convectively  stable regions only}  \\
\hline
\fontsize{8}{5} \selectfont{Propagation cavity strongly dependent on $l$} &
\fontsize{8}{5}\selectfont {Propagation cavity
largely independent of  $l$}  \\
\hline
\fontsize{8}{5} \selectfont{Frequency increases with increasing $n$
  and increasing $l$} &
\fontsize{8}{5}\selectfont {Frequency always $<\max(N_0)$. It increases with increasing $l$,
 and decreases with increasing $|n|$}  \\
\hline
\end{tabular}
\end{table}

\subsection{The surface gravity waves\label{fmode}}
For completeness, in this section I will consider the case of
perturbations obeying  $\delta p=0$, hence, $\Psi=0$.
Going back to the system of equations (\ref{amp_cow}), and 
considering, as before, that locally the oscillations can be treated
as in a plane-parallel layer under constant gravity, we find:
\begin{eqnarray}
\frac{{\rm d}\xi_r}{{\rm
    d}r}-\frac{g_0\kappa_h^2}{\omega^2}\xi_r
+\frac{1}{\rho_0c_0^2}\left(1-\frac{c_0^2\kappa_h^2}{\omega^2}\right)\delta
p &
= & 0 \,,\nonumber \\
\frac{{\rm d}\delta p}{{\rm d}r}+\frac{g_0\kappa_h^2}{\omega^2}\delta
p-\rho_0
g_0\left(\frac{\omega^2}{g_0}-\frac{g_0\kappa_h^2}{\omega^2}\right)\xi_r
&=& 0\,.
\label{amp_dp0}
\end{eqnarray}
For $\delta p=0$,  the system above is satisfied by an
exponential solution of the type $\xi_r=\exp \left[\kappa_h\left(
    r-r_0\right)\right]$, where $\kappa_h=\omega^2/g_0$.  Here,  $r_0$ is a fiducial depth in the
vicinity of which the plane-parallel approximation is being made and
 $g_0$ is the gravitational acceleration at $r=r_0$.
Thus, the depth-dependent amplitude of the solution in this case
decays exponentially with depth and the dispersion relation is $\omega=\sqrt{g_0\kappa_h}$. Moreover, under te approximations
considered here, it is independent of
the stratification of the star. Since the characteristic
scale of the amplitude decay is $\kappa_h^{-1}$,  the plane-parallel
approximation is particularly adequate when $\kappa_h$,
hence the degree $l$ of the mode, is not too small and the mode is concentrated near the surface of the star. 


This solution can be identified as a surface gravity wave, similar to a wave propagating at the surface of a deep
ocean.  Since $\delta p=0$, $\nabla\cdot\boldsymbol{\xi}=0$ and the
fluid is not compressed during the perturbation.

\section{Numerical solutions to the pulsation equations\label{numerical}}
The full solutions to the pulsation equations and
associated boundary conditions must be computed numerically. In this
section, I briefly discuss the range of solutions obtained for a model of the Sun and
 discuss also a specific example for a star in a
different evolutionary state.

\begin{figure}[t]
\centering
\hspace{-0.45cm}
\includegraphics[scale=0.62]{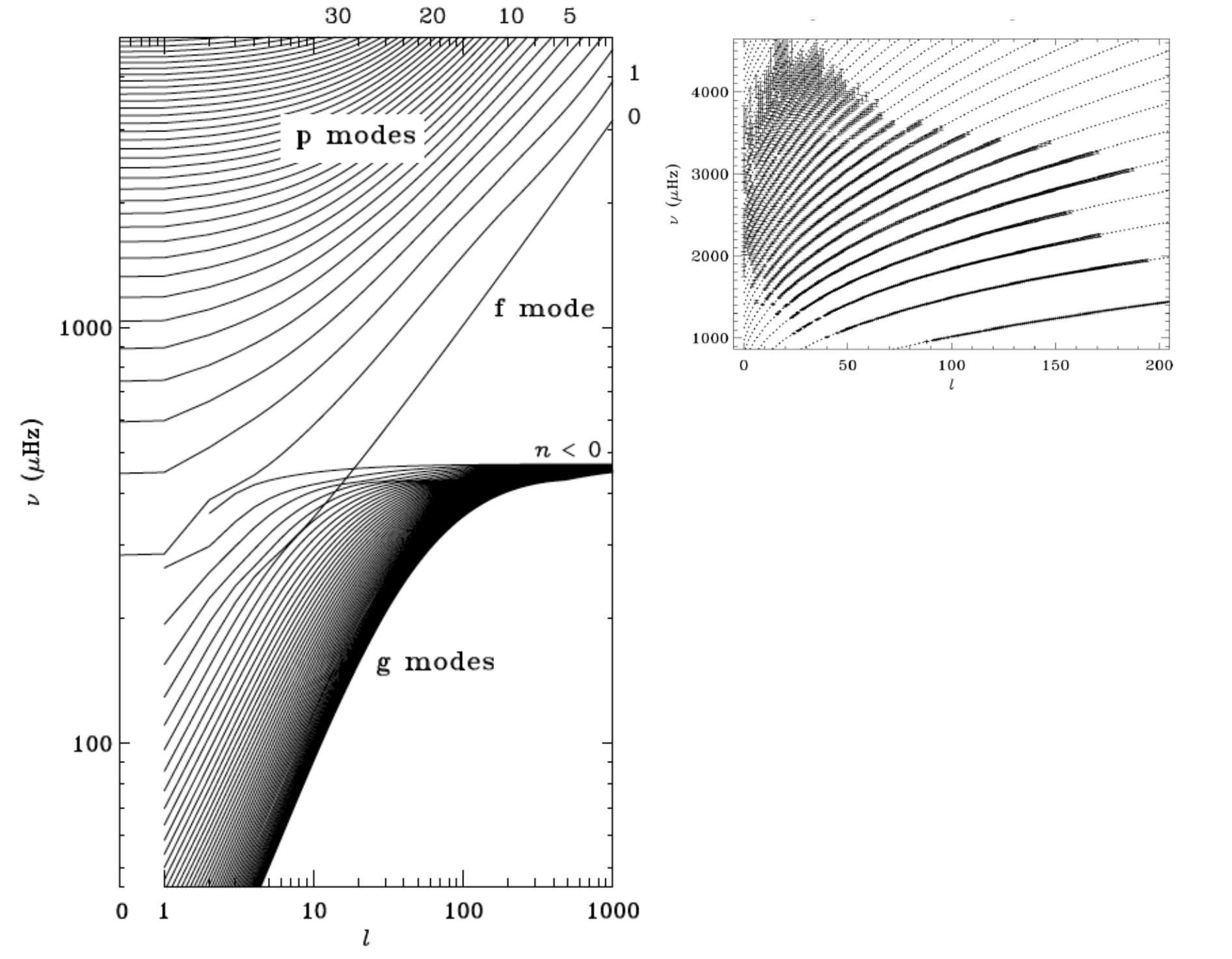} 

%
%
\caption{\textit{Left:} Cyclic frequencies computed for a model of
  the Sun, as function of mode degree, $l$. The
discrete eigenvalues for each radial order have been joined by
continuous lines with a few examples of the radial order identified on
the right-hand side and top of the figure. \textit{Right:} Frequencies of the Sun derived from 144 days of observations with the
instrument MDI on board the \textit{SOHO} spacecraft. The depicted error bars correspond
to 1000$\sigma$. The dotted lines show the
model results, for comparison. Figure adapted
from \citet{jcd08c,aerts10}.}
\label{eigen}       
\end{figure}

The left panel of Fig.~\ref{eigen}, adapted from \citet{aerts10}, shows the cyclic
frequencies (=$\omega/2\pi)$ computed with the Aarhus adiabatic
oscillation package \citep[\textsc{adipls};][]{JCD08b} for a solar model obtained
with the Aarhus STellar Evolution Code \citep[\textsc{astec};][]{JCD08a}. The
discrete eigenvalues for each radial order have been joined by
continuous lines with a few examples of the radial order identified on
the right-hand side and top of the figure. Three families of
solutions are identified in the figure: (i) The acoustic (or p-)
modes, at higher frequencies; (ii) the gravity (or g-) modes, at
lower frequencies, and; (iii) the f-mode, at intermediate frequencies.
These correspond to the cases discussed in the previous section, based
on the analysis under the Cowling approximation.
For comparison, the eigenfrequencies of the real Sun, derived from data
acquired with the instrument MDI on board of the \textit{SOHO} spacecraft are shown on the
right panel of the same figure, where the model results are
overplotted as dotted curves. Only the p- and f-modes are seen in the
real data. In fact, the observation of gravity modes in the Sun has
been a long-standing goal, but despite all efforts and some claims of
detection of signatures of g-modes and, possibly, individual g-modes in the Sun \citep{garcia07, garcia10},
the matter is still not settled \citep{appourchaux10}.

A closer look at the different families of solutions displayed in
Fig.~\ref{eigen} shows that the behaviour of the solutions with
changing radial order and mode degree also follows what was
found from the analysis in Sect.~\ref{trapping}. For the p-modes, at
fixed degree, $l$, we can see an increase of the frequency with radial
order. Likewise, following a single line of fixed radial roder, $n$,
we also see an increase of the frequency with increasing degree. Both
of these dependencies were expected from the dispersion relation for
acoustic waves derived earlier. Regarding the g-modes, we can identify
an upper bound to the frequency, which corresponds to the
maximum value of the buoyancy frequency in that model. In addition, I
note that there are no results for spherically symmetric modes, $l=0$,
as expected given that the displacement associated with these waves
can never be purely radial (always involves a horizontal
displacement of the fluid). Moreover, it is seen that at fixed degree the
frequency decreases with increasing absolute value of the radial order,
$|n|$, while at fixed radial order, the frequency increases with
increasing degree. I also note that the fact that the lines of g-mode
solutions are very close to each other implies that at fixed frequency
the radial order increases very rapidly with increasing mode degree. 
As for the case of p-modes, the numerical results discussed here are in agreement
with the dispersion relation for g-modes derived in
Sect.~\ref{trapping}. Finally, the f-mode eigenfrequencies for moderate
to high mode degrees (the ones for which the plane-parallel approximation used
in Sect.~\ref{trapping} is adequate), are found between the p-mode and
g-mode eigenfrequencies. Despite the resemblance of the high degree
f-mode solutions and the p-mode solutions I wish to recall that these
modes obey $\delta p=0$. Thus, the perturbation takes place without compression or refraction of the
fluid, reminding us that f-modes are not acoustic waves.
\begin{figure}[t]
\hspace{-0.8cm}\includegraphics[scale=0.41]{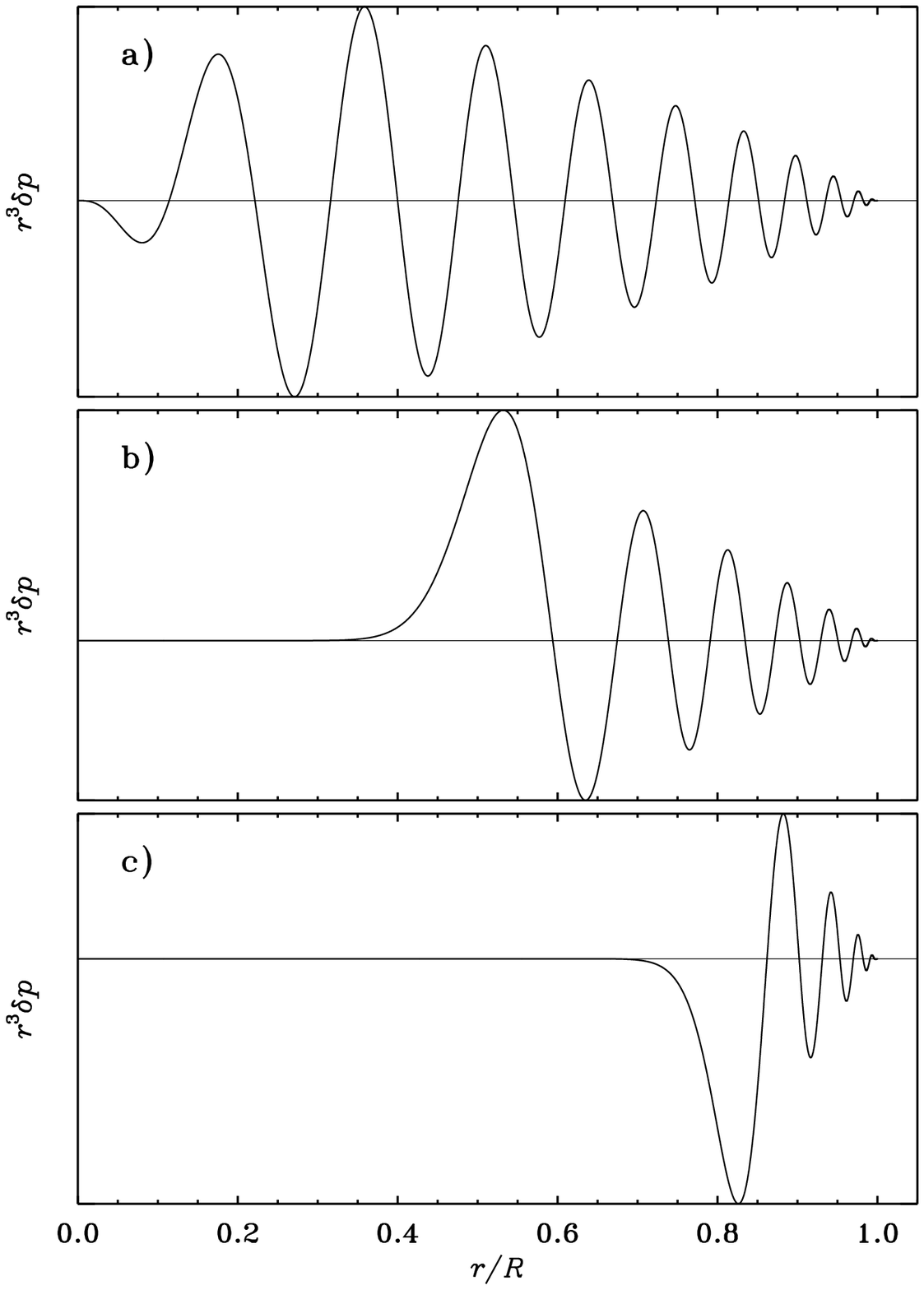} 
\hspace{-0.8cm}\includegraphics[scale=0.41]{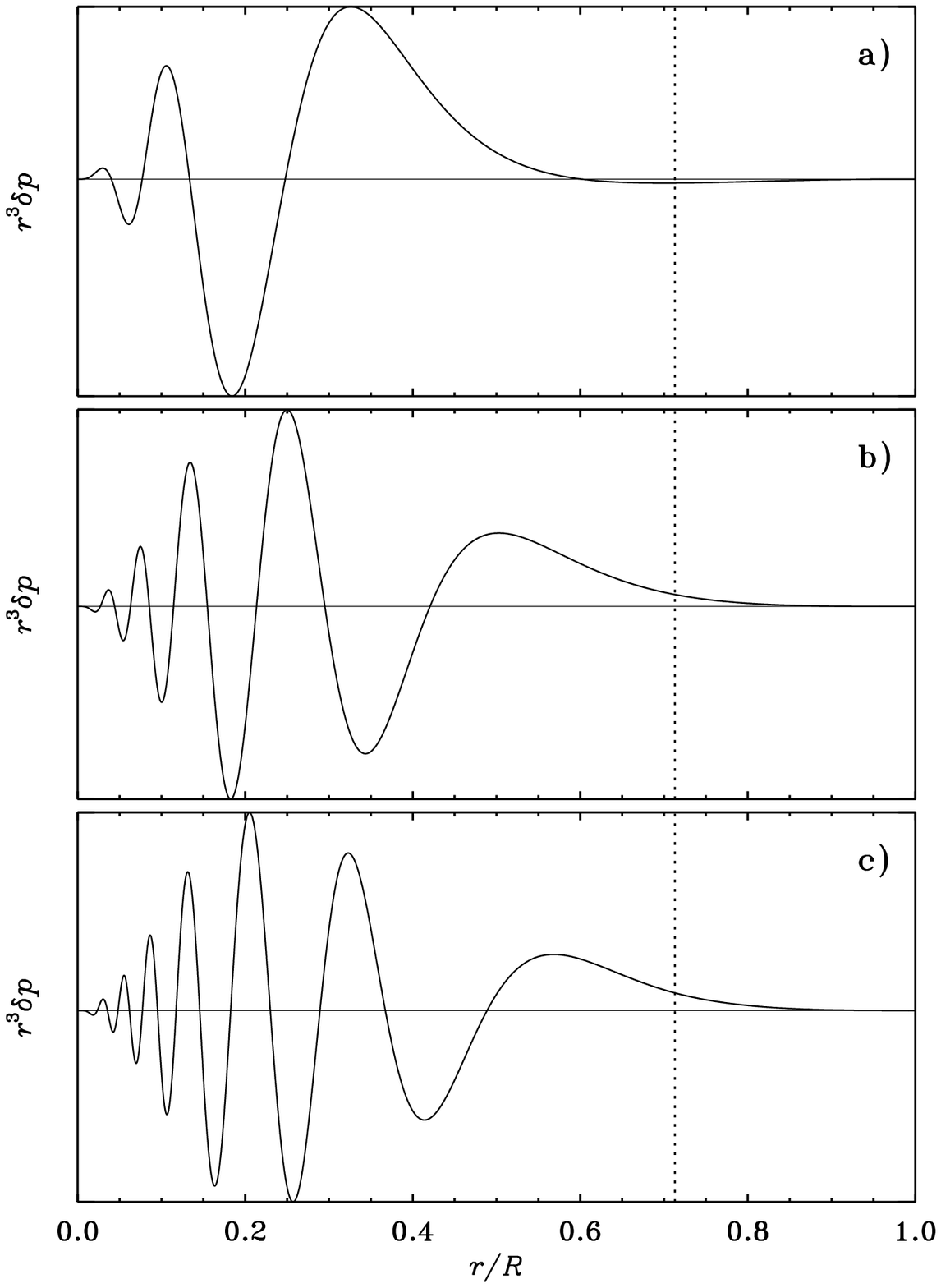} 

%
%
\caption{Normalized eigenfunctions for a model of the Sun, as function of
  fractional  radius. The chosen eigenfunction, $r^3\delta p$, has the dimensions of energy and is normalized to its
  maximum value. \textit{Left:} Results for 3 p-modes:
  a) ($l=0, n=21, \nu =3038.0\mu$Hz), b) ($l=20, n=14, \nu =2939.2\mu$Hz),
  and c) ($l=60, n=9, \nu =3043.2\mu$Hz). \textit{Right:} Results for 3 g-modes:
  a) ($l=1, n=-5, \nu =109.2\mu$Hz), b) ($l=2, n=-10, \nu =102.6\mu$Hz),
  and c) ($l=3, n=-14, \nu =104.1\mu$Hz). The dotted line marks the
  base of the convective envelope. (Figure
courtesy of J\o rgen Christensen-Dalsgaard.)}
\label{eigenfunctionsS}       
\end{figure}

\begin{figure}[t]
\sidecaption
\hspace{0.cm}\includegraphics[scale=0.162]{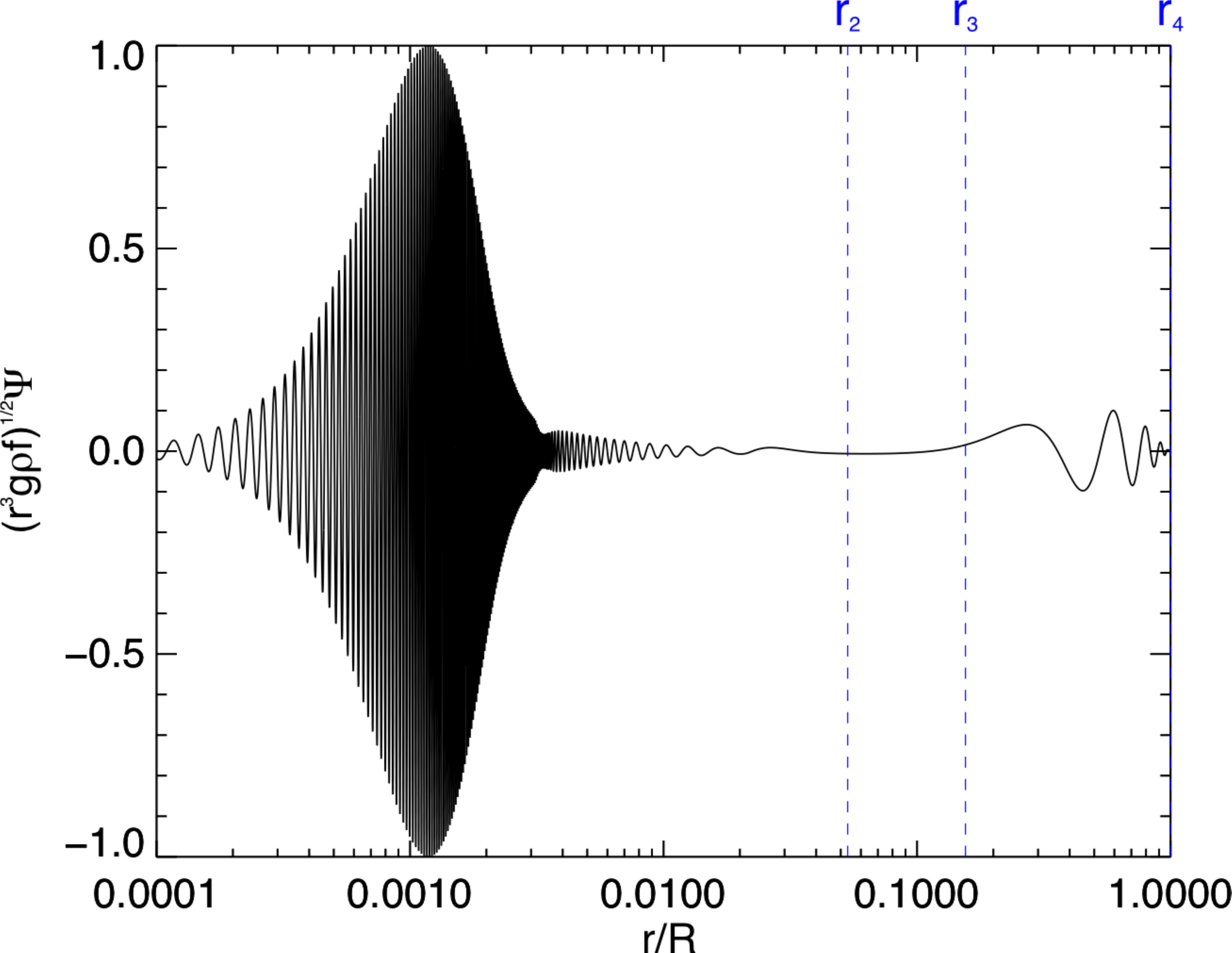} 
%
%
\caption{Same as Fig.~\ref{eigenfunctionsS}, for a
dipole mode with frequency $\nu = 51.20\mu$Hz, in a model of a star in
the red-giant branch. The vertical, blue dashed lines show the
position of the two turning points
bounding the evanescent region, namely, $r_2$ and $r_3$. The outermost turning point, $r_4$, is also shown,
while the innermost turning point, $r_1$, is outside the plotted range. The g-mode
cavity is located between the unseen $r_1$ and $r_2$, and the p-mode cavity is located
between $r_3$ and $r_4$. Figure adapted from \citet{cunha15}.}
\label{eigenfunctionRG}       
\end{figure}

The trapping of the modes discussed in Sect.~\ref{trapping} can also be
verified by inspection of the numerical eigenfunctions. In
Fig.~\ref{eigenfunctionsS} three examples of p-mode (left panel)
and g-mode (right panel) eigenfunctions are shown, for the solar model discussed
above. The quantity plotted,  $r^3\delta p$,  has the dimensions of energy and, in each case, it is
normalized to its maximum value. Inspection of the p-mode solutions shows that the spherically
symmetric pulsation ($l=0$; top panel) propagates from the centre to
the stellar surface. One may then expected it to carry average information about
the entire star. In contrast, a
p-mode of similar frequency but much higher degree ($l=60$; bottom panel) has its
energy concentrated in the outer layers of the star. This was
expected from the analysis performed in Sect.~\ref{trapping}, where it was
found that the propagation cavity of p-modes becomes shallower as the
mode degree increases. 
Looking now at the right panel we see that, in contrast with the p-modes,
the energy of the g-modes is concentrated towards the
innermost layers of the star.  The modes are trapped below the
convective envelope (marked by the vertical dotted line), in a
cavity that is mostly independent of mode degree.  The modes shown
have similar frequency, so as anticipated from Fig.~\ref{eigen}, even
a small increase in mode degree results is a significant increase in
the radial order, as seen from the increase in the number of nodes
when comparing the upper and lower panels on the right-hand side.

The regions where modes propagate depend directly on the stellar
structure and, hence, are different for stars of different masses or
different evolutionary states. In Sect.~\ref{solutions} I have
pointed out that as the star evolves beyond the main sequence, and
the core contracts, the buoyancy frequency increases significantly
towards the centre. This results in the appearance of mixed modes,
i.e., modes that are maintained by gravity acting on density perturbations in the
deep interior, and by the gradient of the pressure perturbation in the outer
layers. An example of such a mode computed for the model shown in
Fig.~\ref{freqRG} is shown in
Fig.~\ref{eigenfunctionRG}.  Two mode
cavities can be identified, in the inner and outer layers of the star,
respectively, separated by an evanescent region where the solution is
not oscillatory.  Despite the latter, coupling does exist between the
two cavities and in a general case the solution is different from what would
be found if the two cavities were  considered independently of each other.

\section{Discussion\label{conclusions}}

As mentioned from the outset, a number of aspects of the theory of
stellar pulsations had to be left out of these notes in the interest
of space. In what follows, I identify issues that I find
particularly important and that are discussed in detail in the books
and lecture notes mentioned in the introduction.

To start with, it should be pointed out that it is possible to
perform an asymptotic analysis of the second-order pulsation equations
derived under the Cowling approximation to find approximate
eigenvalues and eigenfunctions for modes of high radial orders and low
degree. That has been performed in different ways by different authors \citep{vandakurov68,tassoul80,
unno89,gough93}. In the case of p-modes, the eigenfrequencies in this
limit are found to be well approximated by
\begin{equation}
\nu_{nl}  \simeq
\left( n + {l \over 2 }+ {1 \over 4 }+ \alpha \right) \Delta \nu_0
- [A l(l+1) - \delta ] {\Delta \nu_0^2  \over \nu_{nl}} \,,
\label{eq:pasymp}
\end{equation}
where
\begin{equation}
\Delta \nu_0 = \left(2 \int_0^R {\dd r \over c} \right)^{-1}
\label{eq:lsep}
\end{equation}
is the inverse sound travel time across a stellar diameter, and
\begin{equation}
A = {1 \over 4 \pi^2 \Delta \nu_0}
\left[ {c(R) \over R }- \int_0^R {\dd c  \over \dd r} {\dd r \over r }\right]\,.
\label{eq:pcorr}
\end{equation}
Here, $\alpha$ is a slowly varying function of frequency
determined by the reflection properties near the surface
and $\delta$ is a small correction term predominantly related to the
near-surface region.
To leading order, Eq.~(\ref{eq:pasymp}) predicts 
the uniform spacing between frequencies of consecutive modes of the same degree,
corresponding to the large frequency separation, $\Delta\nu$,
observed in Fig.~\ref{sun}. Also, we see that modes of odd degree fall halfway between
modes of even degree.
The first term in Eq.~(\ref{eq:pasymp}) cancels out when subtracting the frequencies of modes of consecutive orders and degrees
  differing by two. That combination then gives the small frequency
  separation, $\delta\nu$, also seen in Fig.~\ref{sun}, and which is found
  to be
\begin{equation}
\delta \nu_{nl} = 
\nu_{n\,l} - \nu_{n-1 \, l+2} \simeq
- ( 4 l + 6 ) {\Delta \nu_0 \over 4 \pi^2 \nu_{nl}}
\int_0^R {\dd c \over \dd r }{\dd r \over r } \, ,
\label{eq:smallsep}
\end{equation}
where the small term $c(R)$ in
Eq.~(\ref{eq:pcorr}) has been neglected. From Eq.~(\ref{eq:smallsep}) we see that the small
separation is particularly sensitive to the innermost layers, as a
result of the $r^{-1}$ dependence of the integrand. Other small
separations and ratios of small to large separations are often
considered in asteroseismology. In particular, the ratios \citep{roxburgh03, cunha07} have the advantage that they are
essentially independent of the inadequately modelled
surface layers of the stars. 

In the case of high radial order, low-degree g-modes, the first-order
term of the asymptotic
analysis predicts a uniform spacing in mode period, $\Pi_{nl}$, rather
than in frequency. In this case, we have \citep[e.g.,][]{tassoul80,smeyers07}: 
\begin{equation}
\Pi_{nl} \simeq {\Pi_0 \over \sqrt{l(l+1)}} 
\left(n + \alpha_{l, {\rm  g}} \right)\,,
\label{eq:gasymp}
\end{equation}
where 
\begin{equation}
\Pi_0 = 2 \pi^2 \left( \int_{r_1}^{r_2} N {\dd r \over r} \right)^{-1}\,.
\label{eq:gspacing}
\end{equation}
The phase $\alpha_{l, {\rm g}}$ depends on whether the core of the
star is radiative or convective, depending on mode degree in the first
case but not in the latter. In the case of a radiative core one can
write $\alpha_{l, {\rm g}}=l/2+ \alpha_{\rm g}$, where $\alpha_{\rm g}$ is
 independent  of mode degree. Note, however, that in both cases there is a strong
 dependence on mode degree of the  period spacings between modes of
the same degree and consecutive orders, due to the term
$[l(l+1)]^{-1/2}$ in Eq.~(\ref{eq:gasymp}).  This is in contrast with the case of
p-modes for which the asymptotic large separation is, to first order, independent of
mode degree.

Another important aspect that has been left out of this chapter is
the impact on the oscillation spectrum of rotation and magnetic
fields. A perturbative analysis of the impact of rotation on
pulsations can be found, e.g., in \citet{aerts10}, while the impact of an
internal magnetic field can be
found, e.g., in \citet{gough93}. Non-pertubative analyses of
these phenomena, required in the cases of fast rotation or magnetic
fields that permeate the stellar surface, are discussed
by \citet{lignieres06,reese06} and \citet{cunha00,cunha06}, respectively. 

The ultimate goal of stellar pulsations studies is to infer
information about the physics and dynamics of stellar interiors
from the asteroseismic data. That is commonly achieved through forward
modelling, a method that is intrinsically model-dependent, or, in optimal
cases, through inverse techniques, in which the solutions are not
restricted to those of a set of models.

Finally, I note that no word has been said about the driving of the
pulsations, with the exception of a brief reference, in
Sect.~\ref{intro}, to the fact that modes can be intrinsically stable,
as in solar-like pulsators, or unstable, as in classical
pulsators. This topic is, however, well out of the scope of the present
notes and I, thus, refer the interested reader to the literature
listed in Sect.~\ref{intro}, and references therein.

\begin{acknowledgement}
I am grateful to \^{A}ngela Santos for producing Fig.~\ref{sun} and to J{\o}rgen
Christensen-Dalsgaard for producing Fig.~\ref{eigenfunctionsS}. This work was
supported by Funda\c c\~ao para a Ci\^encia e a Tecnologia (FCT)
through national funds (UID/FIS/04434/2013) and by FEDER through
COMPETE2020 (POCI-01-0145-FEDER-007672) and through the Investigador FCT
Contract No.~IF/00894/2012/CP0150/CT0004. This work has received
funding from EC, under FP7, through the grant agreement FP7-SPACE-2012-312844.
\end{acknowledgement}

\bibliographystyle{apj}
\bibliography{solar-like} 

\begin{thebibliography}{}
\expandafter\ifx\csname natexlab\endcsname\relax\def\natexlab#1{#1}\fi

\bibitem[{{Aerts} {et~al.}(2010){Aerts}, {Christensen-Dalsgaard}, \&
  {Kurtz}}]{aerts10}
{Aerts}, C., {Christensen-Dalsgaard}, J., \& {Kurtz}, D.~W. 2010,
  {Asteroseismology} (Springer)

\bibitem[{{Appourchaux} {et~al.}(2010){Appourchaux}, {Belkacem}, {Broomhall},
  {Chaplin}, {Gough}, {Houdek}, {Provost}, {Baudin}, {Boumier}, {Elsworth},
  {Garc{\'{\i}}a}, {Andersen}, {Finsterle}, {Fr{\"o}hlich}, {Gabriel}, {Grec},
  {Jim{\'e}nez}, {Kosovichev}, {Sekii}, {Toutain}, \&
  {Turck-Chi{\`e}ze}}]{appourchaux10}
{Appourchaux}, T., {Belkacem}, K., {Broomhall}, A.-M., {et~al.} 2010, A\&A
  Rev., 18, 197

\bibitem[{{Baglin} {et~al.}(2006){Baglin}, {Auvergne}, {Barge}, {Deleuil},
  {Catala}, {Michel}, {Weiss}, \& {COROT Team}}]{baglin06}
{Baglin}, A., {Auvergne}, M., {Barge}, P., {et~al.} 2006, in ESA Special
  Publication, Vol. 1306, The CoRoT Mission Pre-Launch Status - Stellar
  Seismology and Planet Finding, ed. M.~{Fridlund}, A.~{Baglin}, J.~{Lochard},
  \& L.~{Conroy}, 33

\bibitem[{{Basu}(2016)}]{basu16}
{Basu}, S. 2016, Living Reviews in Solar Physics, 13, 2

\bibitem[{{Brown} {et~al.}(1991){Brown}, {Gilliland}, {Noyes}, \&
  {Ramsey}}]{Brown91}
{Brown}, T.~M., {Gilliland}, R.~L., {Noyes}, R.~W., \& {Ramsey}, L.~W. 1991,
  APJ, 368, 599

\bibitem[{{Christensen-Dalsgaard}(2008{\natexlab{a}})}]{JCD08b}
{Christensen-Dalsgaard}, J. 2008{\natexlab{a}}, Ap\&SS, 316, 113

\bibitem[{{Christensen-Dalsgaard}(2008{\natexlab{b}})}]{JCD08a}
---. 2008{\natexlab{b}}, Ap\&SS, 316, 13

\bibitem[{{Christensen-Dalsgaard}(2008{\natexlab{c}})}]{jcd08c}
{Christensen-Dalsgaard}, J. 2008{\natexlab{c}}, in IAU Symposium, Vol. 252, The
  Art of Modeling Stars in the 21st Century, ed. L.~{Deng} \& K.~L. {Chan},
  135--147

\bibitem[{{Christensen-Dalsgaard} {et~al.}(1996){Christensen-Dalsgaard},
  {Dappen}, {Ajukov}, {Anderson}, {Antia}, {Basu}, {Baturin}, {Berthomieu},
  {Chaboyer}, {Chitre}, {Cox}, {Demarque}, {Donatowicz}, {Dziembowski},
  {Gabriel}, {Gough}, {Guenther}, {Guzik}, {Harvey}, {Hill}, {Houdek},
  {Iglesias}, {Kosovichev}, {Leibacher}, {Morel}, {Proffitt}, {Provost},
  {Reiter}, {Rhodes}, {Rogers}, {Roxburgh}, {Thompson}, \& {Ulrich}}]{jcd96}
{Christensen-Dalsgaard}, J., {Dappen}, W., {Ajukov}, S.~V., {et~al.} 1996,
  Science, 272, 1286

\bibitem[{{Cunha}(2006)}]{cunha06}
{Cunha}, M.~S. 2006, MNRAS, 365, 153

\bibitem[{{Cunha} \& {Gough}(2000)}]{cunha00}
{Cunha}, M.~S., \& {Gough}, D. 2000, MNRAS, 319, 1020

\bibitem[{{Cunha} \& {Metcalfe}(2007)}]{cunha07}
{Cunha}, M.~S., \& {Metcalfe}, T.~S. 2007, ApJ, 666, 413

\bibitem[{{Cunha} {et~al.}(2015){Cunha}, {Stello}, {Avelino},
  {Christensen-Dalsgaard}, \& {Townsend}}]{cunha15}
{Cunha}, M.~S., {Stello}, D., {Avelino}, P.~P., {Christensen-Dalsgaard}, J., \&
  {Townsend}, R.~H.~D. 2015, ApJ, 805, 127

\bibitem[{{Cunha} {et~al.}(2007){Cunha}, {Aerts}, {Christensen-Dalsgaard},
  {Baglin}, {Bigot}, {Brown}, {Catala}, {Creevey}, {de Souza}, {Eggenberger},
  {Garcia}, {Grundahl}, {Kervella}, {Kurtz}, {Mathias}, {Miglio}, {Monteiro},
  {Perrin}, {Pijpers}, {Pourbaix}, {Quirrenbach}, {Rousselet-Perraut},
  {Teixeira}, {Th{\'e}venin}, \& {Thompson}}]{cunhaetal07}
{Cunha}, M.~S., {Aerts}, C., {Christensen-Dalsgaard}, J., {et~al.} 2007,
  Astronomy and Astrophysics Review, 14, 217

\bibitem[{{Deubner} \& {Gough}(1984)}]{deubnerandgough84}
{Deubner}, F.-L., \& {Gough}, D. 1984, ARA\&A, 22, 593

\bibitem[{{Domingo} {et~al.}(1995){Domingo}, {Fleck}, \& {Poland}}]{domingo95}
{Domingo}, V., {Fleck}, B., \& {Poland}, A.~I. 1995, Solar Physics, 162, 1

\bibitem[{{Fr{\"o}hlich} {et~al.}(1995){Fr{\"o}hlich}, {Romero}, {Roth},
  {Wehrli}, {Andersen}, {Appourchaux}, {Domingo}, {Telljohann}, {Berthomieu},
  {Delache}, {Provost}, {Toutain}, {Crommelynck}, {Chevalier}, {Fichot},
  {D{\"a}ppen}, {Gough}, {Hoeksema}, {Jim{\'e}nez}, {G{\'o}mez}, {Herreros},
  {Cort{\'e}s}, {Jones}, {Pap}, \& {Willson}}]{frohlich95}
{Fr{\"o}hlich}, C., {Romero}, J., {Roth}, H., {et~al.} 1995, Solar Physics,
  162, 101

\bibitem[{{Garc{\'{\i}}a}(2010)}]{garcia10}
{Garc{\'{\i}}a}, R.~A. 2010, Highlights of Astronomy, 15, 345

\bibitem[{{Garc{\'{\i}}a} {et~al.}(2007){Garc{\'{\i}}a}, {Turck-Chi{\`e}ze},
  {Jim{\'e}nez-Reyes}, {Ballot}, {Pall{\'e}}, {Eff-Darwich}, {Mathur}, \&
  {Provost}}]{garcia07}
{Garc{\'{\i}}a}, R.~A., {Turck-Chi{\`e}ze}, S., {Jim{\'e}nez-Reyes}, S.~J.,
  {et~al.} 2007, Science, 316, 1591

\bibitem[{{Gilliland} {et~al.}(2010){Gilliland}, {Brown},
  {Christensen-Dalsgaard}, {Kjeldsen}, {Aerts}, {Appourchaux}, {Basu},
  {Bedding}, {Chaplin}, {Cunha}, {De Cat}, {De Ridder}, {Guzik}, {Handler},
  {Kawaler}, {Kiss}, {Kolenberg}, {Kurtz}, {Metcalfe}, {Monteiro}, {Szab{\'o}},
  {Arentoft}, {Balona}, {Debosscher}, {Elsworth}, {Quirion}, {Stello},
  {Su{\'a}rez}, {Borucki}, {Jenkins}, {Koch}, {Kondo}, {Latham}, {Rowe}, \&
  {Steffen}}]{guilliland10}
{Gilliland}, R.~L., {Brown}, T.~M., {Christensen-Dalsgaard}, J., {et~al.} 2010,
  PASP, 122, 131

\bibitem[{{Gough}(1993)}]{gough93}
{Gough}, D.~O. 1993, in Astrophysical Fluid Dynamics - Les Houches 1987, ed.
  J.-P. {Zahn} \& J.~{Zinn-Justin}, 399--560

\bibitem[{{Jim{\'e}nez} {et~al.}(2002){Jim{\'e}nez}, {Roca Cort{\'e}s}, \&
  {Jim{\'e}nez-Reyes}}]{jimenez02}
{Jim{\'e}nez}, A., {Roca Cort{\'e}s}, T., \& {Jim{\'e}nez-Reyes}, S.~J. 2002,
  Solar Physics, 209, 247

\bibitem[{{Kjeldsen} \& {Bedding}(1995)}]{kjeldsen95}
{Kjeldsen}, H., \& {Bedding}, T.~R. 1995, A\&A, 293, 87

\bibitem[{{Koch} {et~al.}(2010){Koch}, {Borucki}, {Basri}, {Batalha}, {Brown},
  {Caldwell}, {Christensen-Dalsgaard}, {Cochran}, {DeVore}, {Dunham},
  {Gautier}, {Geary}, {Gilliland}, {Gould}, {Jenkins}, {Kondo}, {Latham},
  {Lissauer}, {Marcy}, {Monet}, {Sasselov}, {Boss}, {Brownlee}, {Caldwell},
  {Dupree}, {Howell}, {Kjeldsen}, {Meibom}, {Morrison}, {Owen}, {Reitsema},
  {Tarter}, {Bryson}, {Dotson}, {Gazis}, {Haas}, {Kolodziejczak}, {Rowe}, {Van
  Cleve}, {Allen}, {Chandrasekaran}, {Clarke}, {Li}, {Quintana}, {Tenenbaum},
  {Twicken}, \& {Wu}}]{koch10}
{Koch}, D.~G., {Borucki}, W.~J., {Basri}, G., {et~al.} 2010, ApJL, 713, L79

\bibitem[{{Ligni{\`e}res} {et~al.}(2006){Ligni{\`e}res}, {Rieutord}, \&
  {Reese}}]{lignieres06}
{Ligni{\`e}res}, F., {Rieutord}, M., \& {Reese}, D. 2006, A\&A, 455, 607

\bibitem[{{Reese} {et~al.}(2006){Reese}, {Ligni{\`e}res}, \&
  {Rieutord}}]{reese06}
{Reese}, D., {Ligni{\`e}res}, F., \& {Rieutord}, M. 2006, A\&A, 455, 621

\bibitem[{{Roxburgh} \& {Vorontsov}(2003)}]{roxburgh03}
{Roxburgh}, I.~W., \& {Vorontsov}, S.~V. 2003, A\&A, 411, 215

\bibitem[{{Smeyers} \& {Moya}(2007)}]{smeyers07}
{Smeyers}, P., \& {Moya}, A. 2007, A\&A, 465, 509

\bibitem[{{Takata}(2005)}]{takata05}
{Takata}, M. 2005, PASJ, 57, 375

\bibitem[{{Takata}(2016)}]{takata16a}
---. 2016, PASJ, 68, 109

\bibitem[{{Tassoul}(1980)}]{tassoul80}
{Tassoul}, M. 1980, ApJS, 43, 469

\bibitem[{{Unno} {et~al.}(1989){Unno}, {Osaki}, {Ando}, {Saio}, \&
  {Shibahashi}}]{unno89}
{Unno}, W., {Osaki}, Y., {Ando}, H., {Saio}, H., \& {Shibahashi}, H. 1989,
  {Nonradial oscillations of stars} (University of Tokyo Press)

\bibitem[{{Vandakurov}(1968)}]{vandakurov68}
{Vandakurov}, Y.~V. 1968, \sovast, 11, 630

\bibitem[{{Verner} {et~al.}(2011){Verner}, {Elsworth}, {Chaplin}, {Campante},
  {Corsaro}, {Gaulme}, {Hekker}, {Huber}, {Karoff}, {Mathur}, {Mosser},
  {Appourchaux}, {Ballot}, {Bedding}, {Bonanno}, {Broomhall}, {Garc{\'{\i}}a},
  {Handberg}, {New}, {Stello}, {R{\'e}gulo}, {Roxburgh}, {Salabert}, {White},
  {Caldwell}, {Christiansen}, \& {Fanelli}}]{verner11}
{Verner}, G.~A., {Elsworth}, Y., {Chaplin}, W.~J., {et~al.} 2011, MNRAS, 415,
  3539

\end{thebibliography}

\end{document}